\documentclass[12pt,a4paper]{article}
\usepackage[utf8]{inputenc}
\usepackage{epsf}
\usepackage{latexsym,amssymb,euscript}
\usepackage[dvips]{graphicx}
\usepackage[numbers,sort&compress]{natbib}
\usepackage{amsmath}
\usepackage{nicefrac}
\usepackage{slashed}
\usepackage{booktabs}
\usepackage[linktocpage]{hyperref}
\usepackage{braket}
\usepackage{chngcntr}
\usepackage{bbold}
\usepackage{ragged2e}
\usepackage{multicol}
\usepackage{graphics}
\usepackage{graphicx}
\usepackage{array}
\usepackage{mciteplus}
\usepackage{caption}
\usepackage{subcaption}
\usepackage{pdfpages}
\usepackage[titletoc]{appendix}
\usepackage[normalem]{ulem}
\graphicspath{{./figures/}}
\hypersetup{
 linktocpage = false,
 urlcolor = urlblue,
 colorlinks = true,
 linkcolor = urlblue,
 anchorcolor = citegreen,
 citecolor = urlblue,
 pdfstartview = {XYZ null null 1.25} 
           }
\usepackage[left=2cm, right=2cm]{geometry}
\usepackage{pstricks}
\usepackage{color}
\usepackage{xcolor}
\definecolor{urlblue}{rgb}{0.2,0.4,0.7}
\definecolor{citegreen}{rgb}{0,0.4,0.2}
\definecolor{linkred}{rgb}{0.9,0.2,0.1}
\usepackage{float}
\usepackage{academicons}
\definecolor{orcidlogocol}{HTML}{A6CE39}
\usepackage{changepage}
\usepackage{fancyhdr}
\newcommand{\drv}{{\rm d}}

\newcommand{\LQCD}{\Lambda_{\rm QCD}}

\newcommand{\LL}{{\rm LL/LO}}

\newcommand{\NLLp}{{\rm NLL/NLO^+}}

\newcommand{\HENLOp}{{\rm HE}\mbox{-}{\rm NLO^+}}

\newcommand{\DY}{\Delta Y}

\newcommand{\vqTTa}{\langle {\vec q}_T^{\;2} \rangle}

\newcommand{\Jpsi}{J/\psi}

\newcommand{\BCs}{B_c(^1S_0)}
\newcommand{\Bss}{B_c(^3S_1)}

\newcommand{\QXQq}{X_{Q\bar{Q}q\bar{q}}}

\newcommand{\QXcu}{X_{c\bar{c}u\bar{u}}}
\newcommand{\QXcs}{X_{c\bar{c}s\bar{s}}}
\newcommand{\QXbu}{X_{b\bar{b}u\bar{u}}}
\newcommand{\QXbs}{X_{b\bar{b}s\bar{s}}}
\newcommand{\QXbc}{X_{b\bar{b}c\bar{c}}}
\newcommand{\TQQ}{T_{4Q}}
\newcommand{\TQc}{T_{4c}}

\newcommand{{\HFNRevo}}{HF-NRevo}

\newcommand{{\Jethad}}{\tt JETHAD}
\newcommand{{\symJethad}}{\tt symJETHAD}
\newcommand{{\psymJethad}}{\tt (sym)JETHAD}
\newcommand{{\Hell}}{\tt HELL}
\newcommand{{\RadISH}}{\tt RadISH}
\newcommand{{\Pegasus}}{\tt QCD-PEGASUS}
\newcommand{{\HOPPET}}{\tt HOPPET}
\newcommand{{\QCDNUM}}{\tt QCDNUM}
\newcommand{{\APFEL}}{\tt APFEL}
\newcommand{{\APFELpp}}{\tt APFEL++}
\newcommand{{\APFELppp}}{\tt APFEL(++)}
\newcommand{{\EKO}}{\tt EKO}
\newcommand{{\FeynCalc}}{\tt FeynCalc}

\newcommand{\orcidJVB}{\href{https://orcid.org/0000-0002-3917-306X}{\includegraphics[scale=0.1]{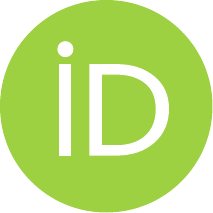}}}

\newcommand{\orcidFGC}{\href{https://orcid.org/0000-0003-3299-2203}{\includegraphics[scale=0.1]{logo-orcid.pdf}}}

\newcommand{\orcidVPG}{\href{https://orcid.org/0000-0003-4943-9973}{\includegraphics[scale=0.1]{logo-orcid.pdf}}}

\renewcommand{\arraystretch}{1.6}

\begin{document}

\begin{titlepage}

{
\begin{adjustwidth}{-1cm}{-1cm}
\begin{center}
  {\LARGE \bf The $b\bar{b}c\bar{c}$ Fully Heavy Tetraquark 
  \\ 
  at the LHC and the FCC:
  \\[0.30cm] 
  Multimodal Fragmentation and Observability}
\end{center}
\end{adjustwidth}
}

\vskip 0.75cm

\centerline{
João~V.~Bulhões,$^{\,1,2\dagger}${\orcidJVB}
Francesco~G.~Celiberto,$^{\,2\ddagger}${\orcidFGC}
and
Victor~P.~Gon\c{c}alves$^{\,1\S}${\orcidVPG}
}

\vskip .4cm

\centerline{${}^1$ {\sl Institute of Physics and Mathematics, Federal University of Pelotas, 354, 96010-900, Pelotas, RS, Brazil}}
\vskip 0.2cm
\centerline{${}^2$ {\sl Universidad de Alcal\'a (UAH), Departamento de Física y Matemáticas,}}
\centerline{\sl E-28805 Alcal\'a de Henares, Madrid, Spain}
\vskip 1.15cm

\begin{abstract}
\vspace{0.25cm}
\headrule 
\vspace{0.50cm}
We investigate the production of the all-heavy mixed-flavor $b\bar b c\bar c$ tetraquark through a multimodal fragmentation setup combining direct-multiquark and scalar-diquark mechanisms.
Standard Dokshitzer--Gribov--Lipatov--Altarelli--Parisi (DGLAP) evolution is applied while preserving threshold consistency through the Heavy-Flavor NonRelativistic evolution (HF-NRevo) scheme.
The resulting fragmentation functions (FFs) are then embedded into hybrid high-energy and collinear factorization at next-to-leading order for semi-inclusive tetraquark-plus-jet production.
We compare the two mechanisms through rapidity-separation distributions at the 14~TeV High-Luminosity LHC (HL-LHC) and at the 100~TeV Future Circular Collider (FCC), finding a clear phenomenological preference for the direct-multiquark channel over the scalar-diquark mode.
Integrated cross sections and expected event yields provide favorable production-level prospects for future searches at the HL-LHC, with substantially enhanced statistics at FCC energies.
These results establish a quantitative connection between microscopic formation mechanisms and the prospective observability of the $b\bar b c\bar c$ tetraquark at present and future hadron colliders.
\vspace{0.50cm} 
\headrule
\vspace{0.75cm}
{
 \setlength{\parindent}{0pt}
 \textsc{Keywords}: \vspace{0.15cm} \\ 
 Exotic matter; 
 Heavy tetraquarks; 
 Heavy flavor; 
 Hadronic structure; 
 High-energy QCD; 
 Fragmentation; 
 Resummation; 
 {\tt TQ4Qhf1.0} FF release
\vspace{0.65cm} 
\headrule
}
\end{abstract}
\vspace{-0.00cm}
\vfill
$^{\dagger}${\it e-mail}:
\href{mailto:joaovitor1729@gmail.com}{joaovitor1729@gmail.com}
\\
$^{\ddagger}${\it e-mail}:
\href{mailto:francesco.celiberto@uah.es}{francesco.celiberto@uah.es}
\\
$^{\S}${\it e-mail}:
\href{mailto:barros@ufpel.edu.br}{barros@ufpel.edu.br}

\end{titlepage}

\tableofcontents
\clearpage

\section{Introduction}
\label{sec:intro}
Heavy-meson production in hadronic collisions at the Large Hadron Collider (LHC) provides a stringent test of perturbative quantum chromodynamics (pQCD) and a key probe of hadronization. Over the past few decades, major experimental efforts have measured heavy-flavor observables with steadily improving precision, and a variety of factorization frameworks have been developed~\cite{Gluck:1977zm,Ellis:1976fj,Bodwin:1994jh,Nayak:2005rt,Ma:2017xno}, particularly for quarkonium production (For a review, see Ref.~\cite{Lansberg:2019adr}). These studies indicate that, at large transverse momentum $p_T$, the dominant quarkonium-production mechanism in hadronic collisions is fragmentation: a high-energy parton (gluon or heavy quark) is produced at short distances and subsequently splits or hadronizes into a quarkonium state. 
In this formalism~\cite{Nayak:2005rt}, the main ingredients for deriving realistic predictions for the production rates of a specific quarkonium state ${\cal{Q}}_i$ are the fragmentation functions (FFs) for the $g \rightarrow {\cal{Q}}_i$ and $Q \rightarrow {\cal{Q}}_i$ transitions, where $Q$ is a heavy quark constituent of ${\cal{Q}}_i$.

In parallel, the past few decades have witnessed tremendous progress in both theoretical and experimental studies of exotic hadrons composed of four or five quarks (For reviews see, {\it e.g.} Refs.~\cite{Liu:2019zoy,Chen:2022asf}). 
In particular, a new resonance denoted $X(6900)$ was reported by the LHCb Collaboration in 2020 in the double-$J/\psi$ invariant-mass spectrum~\cite{LHCb:2020bwg} and later confirmed by the ATLAS~\cite{ATLAS:2023bft} and CMS~\cite{CMS:2023owd} Collaborations. 

More recently, the CMS Collaboration performed an angular analysis of the three all-charm structures $X(6600)$, $X(6900)$, and $X(7100)$, favoring a common $J^{PC}=2^{++}$ assignment~\cite{CMS:2025fpt}. Further experimental evidence was provided by ATLAS through the observation of the $X(6900)$ in the $J/\psi+\psi(2S)$ channel~\cite{ATLAS:2025nsd}, while a subsequent CMS analysis established the three structures and their mutual interference with statistical significances exceeding five standard deviations~\cite{CMS:2026tiu}.

These experimental developments have motivated substantial theoretical efforts to clarify the spectroscopy, internal structure, and production mechanisms of all-charm $c\bar{c}c\bar{c}$ tetraquark candidates.
In this context, several dynamical production mechanisms for all-heavy tetraquarks have been proposed~\cite{Berezhnoy:2011xy,Carvalho:2015nqf,Maciula:2020wri,Zhu:2020xni,Feng:2020riv,Feng:2023agq,Abreu:2023wwg,Celiberto:2024mab,Bai:2024ezn,Belov:2024qyi,Bai:2024flh,Ma:2025ryo,Celiberto:2025dfe,Celiberto:2025ziy,Wang:2025hex,Celiberto:2025vra,Celiberto:2026kks,Wang:2026asl}, largely inspired by approaches developed for quarkonium production. 
Complementary investigations have also explored the two-photon couplings of scalar and tensor all-charm tetraquarks and their production in ultraperipheral heavy-ion collisions~\cite{Goncalves:2021ytq,Biloshytskyi:2022dmo,Chen:2026mrj}.
Of particular interest are Refs.~\cite{Feng:2020riv,Celiberto:2024mab,Bai:2024ezn,Bai:2024flh}, which study inclusive $T_{4c}$ production via fragmentation within the nonrelativistic QCD (NRQCD) factorization framework~\cite{Caswell:1985ui,Thacker:1990bm,Bodwin:1994jh} and predict sizable $T_{4c}$ yields at the LHC. Their results suggest that fragmentation provides a promising avenue for producing fully heavy tetraquark states and for probing their properties.

Motivated by these results, we construct the initial-scale heavy-quark FFs for an all-heavy, mixed-flavor $b\bar b c\bar c$ tetraquark, denoted hereafter as $\QXbc$.
The FFs for the remaining partonic channels, including gluons and light quarks, are then dynamically generated through DGLAP evolution.
Although the $\QXbc$ state has not yet been observed,
its spectroscopy and possible internal configurations
have been investigated within several theoretical
approaches, including potential models, QCD sum rules,
Bethe--Salpeter constructions, and coupled-channel
descriptions (see, {\it e.g.},
Refs.~\cite{Bedolla:2019zwg,Agaev:2024qbh,
Agaev:2024mng,Abreu:2023wwg,Goncalves:2021ytq,
Wu:2024hrv,Ortega:2025lmo,Wang:2025apq,
Liu:2023gla,Wang:2026gch,Likene:2026pga}).
These studies reveal a substantial model dependence
in the predicted spectrum and resonance properties,
further motivating investigations of complementary
tetraquark formation mechanisms.
Complementary studies of exclusive production and
radiative decays have also explored possible
experimental signatures of mixed-flavor all-heavy
tetraquarks~\cite{Liang:2025wbt,Yang:2026rah}.

Our ultimate goal is to provide predictions for $\QXbc$ yields in proton-proton collisions at the 14~TeV High-Luminosity Large Hadron Collider (HL-LHC)~\cite{Apollinari:2015wtw} and the 100~TeV Future Circular Collider (FCC)~\cite{FCC:2025lpp,FCC:2025uan,FCC:2025jtd}, and to assess the prospects for its discovery and for probing its properties. In this paper, we carry out the first two steps toward that objective. 

First, we compute the FFs within two distinct formation mechanisms, allowing us to assess the phenomenological sensitivity to alternative microscopic descriptions of tetraquark hadronization.
In particular, we present the {\tt TQ4Qhf1.0} collinear FF set for the $\QXbc$ state, combining direct-multiquark and sequential scalar-diquark fragmentation modes.

Second, we study the phenomenological implications of $\QXbc$ production via fragmentation by analyzing semi-inclusive $\QXbc$+jet production. We present predictions for the HL-LHC and FCC within hybrid high-energy and collinear factorization at next-to-leading level and beyond ($\NLLp$ HyF). We emphasize rapidity-separation distributions and integrated event yields, which allow us to track how the two microscopic formation mechanisms encoded in the fragmentation formalism propagate to collider-level observables. This comparison provides a direct handle on their relative phenomenological impact and on the prospective observability of the $b\bar b c\bar c$ signal at present and future hadron colliders. 
We postpone to a future publication the study of fully inclusive $\QXbc$ production in $pp$ collisions, without requiring an accompanying tagged jet, together with a comparison of the corresponding predictions with those for other tetraquark states. 

The manuscript is structured as follows.
Section~\ref{sec:FFs} presents the construction of the {\tt TQ4Qhf1.0} FFs for the all-heavy mixed-flavor $b\bar b c\bar c$ tetraquark, together with their standard DGLAP evolution implemented through the threshold-consistent {\HFNRevo} scheme~\cite{Celiberto:2025euy,Celiberto:2024mex}.
Section~\ref{sec:phenomenology} develops the phenomenological analysis of semi-inclusive $\QXbc$ plus jet production at the HL-LHC and FCC, including rapidity-separation distributions, integrated rates, and event-yield estimates for the direct-multiquark and scalar-diquark fragmentation modes.
Finally, Sec.~\ref{sec:conclusions} summarizes our main findings and outlines future developments of the multimodal fragmentation framework toward increasingly realistic and experimentally testable studies of exotic-hadron production.

\section{Heavy-flavor fragmentation to $\QXbc$}
\label{sec:FFs}
The construction of collinear FFs for all-heavy mixed-flavor tetraquarks requires combining perturbative production mechanisms with model-dependent assumptions on the internal structure of the final state.
The {\tt TQ4Qhf1.0} functions address this issue through a multimodal approach, where alternative production mechanisms are incorporated at the level of the initial-scale fragmentation inputs and consistently evolved within a common perturbative framework.

To place the discussion in a broader context, Sec.~\ref{ssec:FFs-intro} summarizes the main theoretical concepts underlying heavy-flavor fragmentation and their extension from conventional hadrons to exotic multiquark systems.
The specific implementation adopted for all-heavy mixed-flavor tetraquarks is then developed through two complementary scenarios.
The direct-fragmentation mechanism is introduced in Sec.~\ref{ssec:FFs-Q-direct}, while the corresponding diquark-based formulation is presented in Sec.~\ref{ssec:FFs-Q-diquark}.
A comparison of the resulting inputs and their impact on the evolved distributions provides the basis of the multimodal {\tt TQ4Qhf1.0} construction.

The final FF set, together with its {\HFNRevo} evolution and the associated perturbative and nonperturbative uncertainty estimates, is discussed in Sec.~\ref{ssec:FFs-TQ4Qhf10}.
The complete construction is designed to provide a flexible platform for investigating the interplay between production dynamics and internal-structure assumptions in all-heavy mixed-flavor tetraquarks.

All symbolic calculations entering the derivation of the {\tt TQ4Qhf1.0} functions are performed with {\symJethad}, a \textsc{Mathematica}~\cite{Mathematica_V14-2} plugin of {\Jethad}~\cite{Celiberto:2020wpk,Celiberto:2022rfj,Celiberto:2023fzz,Celiberto:2024mrq,Celiberto:2024swu,Celiberto:2025csa}, developed for symbolic calculations in hadronic structure and precision QCD.

\subsection{Heavy-flavor fragmentation at a glance}
\label{ssec:FFs-intro}
Heavy-flavor hadronization exhibits a substantially richer structure than its light-flavor counterpart, since the heavy-quark masses entering the lowest Fock configuration remain within the perturbative QCD regime.
Unlike light-hadron FFs, which are entirely nonperturbative, heavy-flavor FFs admit perturbatively calculable initial conditions supplemented by long-distance hadronization effects.

For singly heavy hadrons, including $D$, $B$, and $\Lambda_{c,b}$ states, fragmentation is commonly described through a two-step mechanism~\cite{Cacciari:1996wr,Cacciari:1993mq,Helenius:2023wkn,Generet:2023vte}.
A hard parton $i$, produced at large transverse momentum, first radiates a heavy quark $Q$ through a short-distance process calculable in perturbative QCD owing to the small value of $\alpha_s(m_Q)$.
This contribution defines the short-distance coefficient (SDC) for the transition $[i \to Q]$ and develops on timescales shorter than hadron formation.
The first next-to-leading order (NLO) determinations of these SDCs were derived in Refs.~\cite{Mele:1990yq,Mele:1990cw}, followed by next-to-NLO extensions in Refs.~\cite{Mitov:2006wy,Blumlein:2006rr,Melnikov:2004bm,Mitov:2004du,Biello:2024zti}.
At larger distances, the heavy quark converts into a physical hadron through nonperturbative dynamics modeled either phenomenologically~\cite{Bowler:1981sb,Peterson:1982ak,Colangelo:1992kh} or within effective-theory approaches~\cite{Georgi:1990um,Eichten:1989zv,Neubert:1993mb}.

A consistent zero-mass variable-flavor-number-scheme (ZM-VFNS)~\cite{Mele:1990cw,Cacciari:1993mq} treatment is obtained by evolving these initial conditions through the DGLAP equations, whose coupled evolution kernels encode the scale dependence at the chosen perturbative order.

The same picture extends naturally to quarkonia, although the simultaneous presence of $Q$ and $\bar Q$ inside the lowest Fock state $|Q\bar Q\rangle$ introduces additional dynamical complexity.
Current descriptions are largely based on NRQCD, where production is factorized into perturbative SDCs and nonperturbative long-distance matrix elements (LDMEs) (see Refs.~\cite{Grinstein:1998xb,Kramer:2001hh,Lansberg:2005aw,Lansberg:2019adr}).
In this framework, quarkonium states are represented as a superposition of Fock components organized through a double expansion in $\alpha_s$ and in the relative velocity $v_{\cal Q}$.

The relative importance of the underlying mechanisms depends strongly on the transverse momentum.
At moderate transverse momentum, quarkonia are mainly produced through direct short-distance generation of the $(Q\bar Q)$ pair followed by hadronization.
At sufficiently large transverse momentum, instead, single-parton fragmentation progressively becomes dominant.
The former corresponds to a fixed-flavor-number-scheme (FFNS) higher-power contribution~\cite{Alekhin:2009ni,Kang:2014tta,Boer:2023zit,Celiberto:2025euy,Celiberto:2024mex}, whereas the latter is naturally described within a ZM-VFNS framework driven by DGLAP evolution.

Initial conditions for gluon and heavy-quark fragmentation into $S$-wave vector quarkonia were first computed at LO in Refs.~\cite{Braaten:1993rw,Braaten:1993mp} and later extended to NLO in Ref.~\cite{Zheng:2019gnb}.
These calculations enabled the construction of the first DGLAP-evolved quarkonium FF grids, {\tt ZCW19$^+$}~\cite{Celiberto:2022dyf,Celiberto:2023fzz}, subsequently generalized to $\BCs$ and $\Bss$ mesons in the {\tt ZCFW22} release~\cite{Celiberto:2022keu,Celiberto:2024omj}.
Phenomenological studies based on {\tt ZCFW22} showed that the $\BCs$ yield relative to singly bottomed $B$ mesons stays below the 0.1\% level, consistently with LHCb measurements~\cite{LHCb:2014iah,Celiberto:2024omj}, thereby supporting the fragmentation picture.

These concepts were later extended to exotic hadrons.
NRQCD factorization has been used to interpret double-$\Jpsi$ structures observed at the LHC~\cite{LHCb:2020bwg,ATLAS:2023bft,CMS:2023owd} as compact all-charm tetraquark candidates~\cite{Zhang:2020hoh,Zhu:2020xni}.
In this scenario, $\TQc$ formation proceeds through short-distance production of two charm and two anticharm quarks at scales of order $1/m_c$, followed by nonperturbative binding.
The first NRQCD-based initial condition for gluon fragmentation into color-singlet $S$-wave $\TQc$ states was derived in Ref.~\cite{Feng:2020riv}.

Starting from this result, the first ZM-VFNS FF family for heavy-light tetraquarks, {\tt TQHL1.0}, was proposed in Ref.~\cite{Celiberto:2023rzw} (see Ref.~\cite{Celiberto:2024mrq} for a review).
Later developments produced the {\tt TQ4Q1.1} and {\tt TQHL1.1} sets~\cite{Celiberto:2024beg}, including NRQCD-inspired inputs for the $[Q \to \TQc]$ channel~\cite{Bai:2024ezn}, improved treatments of doubly heavy systems, and bottomoniumlike configurations.
These FFs were subsequently employed in studies of all-heavy tetraquark production~\cite{Celiberto:2025ziy,Celiberto:2025dfe,Celiberto:2026zed}, indirect searches through electroweak and Higgs decays~\cite{Ma:2025ryo}, forward-rapidity analyses sensitive to intrinsic charm~\cite{Celiberto:2025vra}, and dedicated investigations of gluon-induced channels~\cite{Nakhaei:2025zty}.

More recently, the {\tt NRFF1.0} sets~\cite{Celiberto:2025euy} delivered a systematic determination of pseudoscalar quarkonium FFs based on consistent NRQCD inputs, while {\tt TQ4Q2.0} further developed the NRQCD-based construction for all-heavy tetraquark fragmentation through the inclusion of additional production channels and an improved treatment of theoretical uncertainties~\cite{Celiberto:2026kks}.
Within the same program, first FFs for all-charm pentaquarks and rare $\Omega$ baryons were developed in Refs.~\cite{Celiberto:2025ipt,Celiberto:2026rdk,Celiberto:2025ogy,Celiberto:2026qiz}.

\subsection{Direct fragmentation}
\label{ssec:FFs-Q-direct}
The initial-scale modeling of constituent heavy-quark fragmentation into a color-singlet $S$-wave tetraquark $\QXbc$ follows the theoretical construction introduced in Ref.~\cite{Nejad:2021mmp}.
This approach is formulated within a Suzuki-inspired model~\cite{Suzuki:1977km,Suzuki:1985up,Amiri:1986zv}, developed to describe spin-dependent fragmentation processes while preserving an explicit transverse-momentum dependence.
The corresponding collinear limit is recovered by neglecting the relative motion of the constituent quarks inside the bound state~\cite{Lepage:1980fj,Brodsky:1985cr,Amiri:1986zv}.
Calculations presented in Refs.~\cite{Suzuki:1977km,Suzuki:1985up,Amiri:1986zv,Nejad:2021mmp} later provided the theoretical basis for the construction of the initial-scale $[c \to \QXcu]$ and $[b \to \QXbs]$ fragmentation channels implemented in the pioneering {\tt TQHL1.0} set~\cite{Celiberto:2023rzw}.
The subsequent release of the $[c \to \QXcs]$ and $[b \to \QXbu]$ {\tt TQHL1.0} functions followed in Ref.~\cite{Celiberto:2024mrq}.
In the present work, the same general framework is adopted and extended to the direct heavy-quark fragmentation mechanism into the fully heavy tetraquark $\QXbc$.

\begin{figure*}[!t]
\centering
\includegraphics[width=0.475\textwidth]{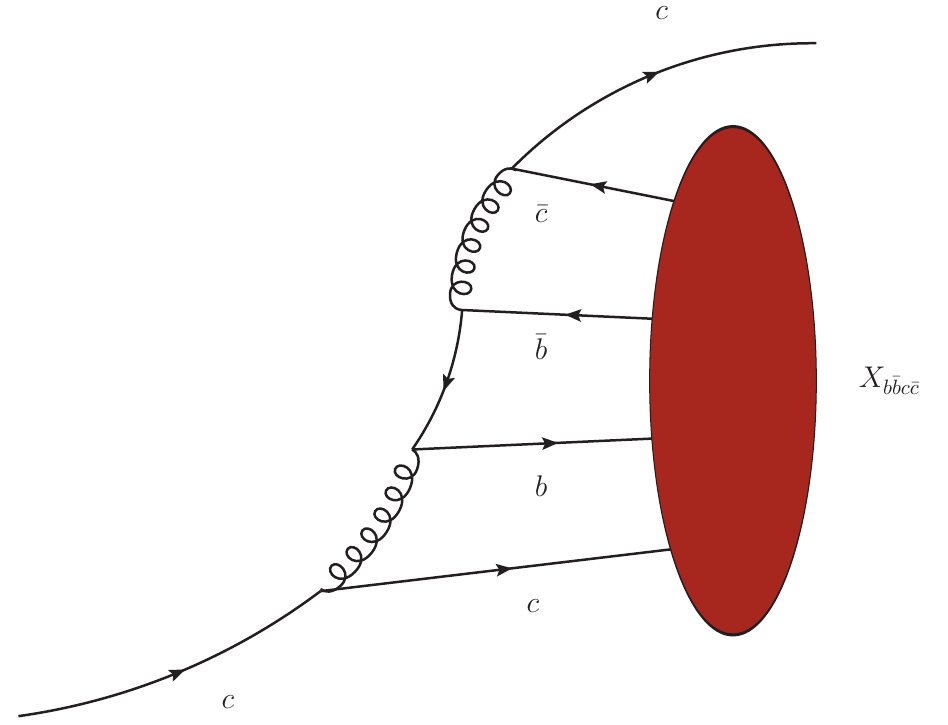}
\hspace{0.00cm}
\includegraphics[width=0.475\textwidth]{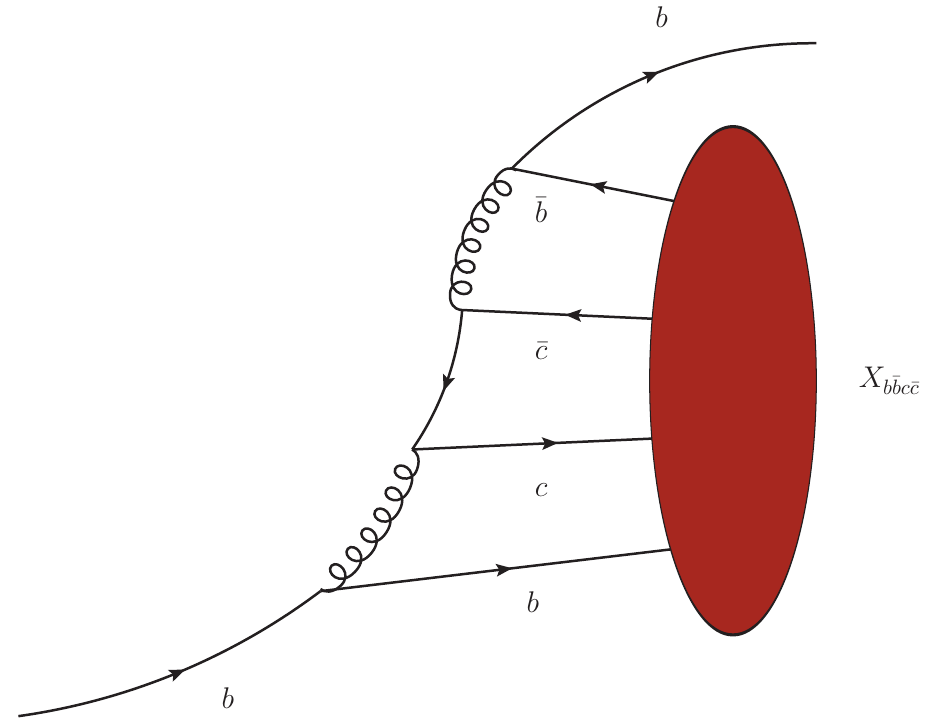}

\caption{Representative LO diagrams for the collinear fragmentation of a constituent charm (left) or bottom (right) quark into a color-singlet $\QXbc$ tetraquark in the direct fragmentation channel.
The firebrick-shaded blob denotes the nonperturbative hadronization component entering the corresponding FF.
Feynman diagrams were produced with {\tt JaxoDraw 2.0}~\cite{Binosi:2008ig}.}
\label{fig:Xbc_FF_drt}
\end{figure*}

The proposed mechanism closely resembles the NRQCD factorization picture, where the heavy-quark pair is produced perturbatively, while the subsequent bound-state formation is encoded in the corresponding LDMEs.
Analogously, within the present framework, the $\QXbc$ system originates from perturbative above-threshold splittings of the outgoing heavy (anti)quark, as schematically shown in Fig.~\ref{fig:Xbc_FF_drt}, where the left and right panels correspond to the charm- and bottom-initiated fragmentation channels, respectively.
The resulting production amplitude is subsequently convoluted with a bound-state wave function describing the nonperturbative tetraquark hadronization dynamics, according to the Suzuki formalism.

Here and in the following, $Q$ denotes the fragmenting heavy quark, while $\tilde{Q}$ labels the second heavy constituent entering the tetraquark bound state. 
Therefore, for the $[c \to \QXbc]$ channel one has $(Q,\tilde{Q})=(c,b)$, whereas for the $[b \to \QXbc]$ transition one has $(Q,\tilde{Q})=(b,c)$.
Under the assumption of complete symmetry between heavy-quark and heavy-antiquark fragmentation channels, the explicit expression of the initial-scale direct FF for the $[Q \to \QXbc]$ transition reads
\begin{equation}
 \label{Xbc_FF_initial-scale_Q_drt}
D_{Q,\,{\rm [direct]}}^{\QXbc}(z,\mu_{F,0}^{(Q)}) \,=\,
{\mathcal N^{(Q)}_{\rm [direct]}} \,
 \frac{(1 - z)^5}{\left[\Xi^{(Q,\tilde{Q})}_{\rm [direct]}(z)\right]^2}
 \sum\limits_{l=0}^3 \, z^{2(l+2)} \rho_{l,\,{\rm [direct]}}^{(Q,\tilde{Q})}(z) 
 \left\{ \frac{\vqTTa}{m_Q^2} \right\}^{\!l}
 \,,
\end{equation}
with
\begin{equation}
 \label{Xbc_FF_initial-scale_Q_N_drt}
 {\mathcal N^{(Q)}_{\rm [direct]}} \, = \, \left\{ 128 \, \pi^2 \, f^{(X)}_{\rm [direct]} \, C_F \big[ \alpha_s(2\mu_X+m_Q) \big]^2 \right\}^2
 \,,
\end{equation}
and $\mu_X = m_Q \,+\, m_{\tilde{Q}}$.
Here, $f^{(X)}_{\rm [direct]}$ denotes the effective nonperturbative normalization parameter controlling the compact tetraquark wave function within the Suzuki-inspired fragmentation model. 
Its phenomenological determination and physical interpretation will be discussed below (see Sec.~\ref{ssec:FFs-Q-diquark}). 
Then, $C_F = (N_c^2-1)/(2N_c)$ is the Casimir factor associated with gluon emission from a quark, while $m_c = 1.5$~GeV and $m_b = 4.9$~GeV denote the charm- and bottom-quark masses, respectively.
One defines
\begin{equation}
 \label{Xbc_FF_initial-scale_Q_Xi_drt}
 \Xi^{(Q,\tilde{Q})}_{\rm [direct]}(z) \,=\,
 \frac{{\mathcal F}_{\rm [direct]}^{(Q,\tilde{Q})}(z)}
 {\mu_X^2 \, m_Q^5 \, m_{\tilde{Q}}}
 \;,
\end{equation}
and 
\begin{equation}
\begin{split}
 \label{Xbc_FF_initial-scale_Q_F_drt}
 {\mathcal F}_{\rm [direct]}^{(Q,\tilde{Q})}(z) &\,=\, \left\{ [ (2-z)\,m_Q + 2\,m_{\tilde{Q}} ]^2 + z^2 \vqTTa \right\} \\[0.10cm]
 \,&\times\,
 \left\{ [ (2-z)\,m_Q + (1-z)\,m_{\tilde{Q}} ]^2 + z^2 \vqTTa \right\}^{2} \\[0.15cm]
 \,&\times\,
 \left\{ 4 (1-z) \mu_X^2 + z^2 (m_Q^2 + \vqTTa) \right\}
 \,.
\end{split}
\end{equation}
The definition of $\Xi_{\rm [direct]}^{(Q,\tilde Q)}$ is chosen so that all coefficient functions $\rho_{l,\rm [direct]}^{(Q,\tilde Q)}$ are dimensionless.
The expressions for the $\rho_{l,\,{\rm [direct]}}^{(Q,\tilde{Q})}(z)$ coefficients in Eq.~\eqref{Xbc_FF_initial-scale_Q_drt} are
\begin{equation}
\begin{split}
 \label{Xbc_FF_initial-scale_Q_rho_0_drt}
 \rho_{0,\,{\rm [direct]}}^{(Q,\tilde{Q})}(z) &\,=\,
 32\, {\cal R}_{\tilde{Q}/Q}^6\, (2 - 3 z + z^3) 
 -
 64\, {\cal R}_{\tilde{Q}/Q}^5\, (6 - 10 z + 3 z^2 + z^3) \\[0.15cm] 
 &\,+\, 8\, {\cal R}_{\tilde{Q}/Q}^4\, (-120 + 236 z - 158 z^2 
 +
 42 z^3 - 6 z^4 + z^5) \\[0.15cm]
 &\,+\, 16\, {\cal R}_{\tilde{Q}/Q}^3\, (-80 + 192 z - 196 z^2 
 +
 98 z^3 - 24 z^4 + 3 z^5) \\[0.15cm] 
 &\,-\, 2\, {\cal R}_{\tilde{Q}/Q}^2\, (480 - 1424 z + 1872 z^2 
 -
 1232 z^3 + 422 z^4 - 71 z^5 + 6 z^6) \\[0.15cm]
 &\,-\, 4\, {\cal R}_{\tilde{Q}/Q}\, (96 - 352 z + 544 z^2 
 -
 424 z^3 + 182 z^4 - 40 z^5 + 3 z^6) \\[0.15cm]
 &\,-\, (4 - 2 z + z^2) (4 - 8 z + 3 z^2)^2 \;,
\end{split}
\end{equation}
\begin{equation}
\begin{split}
 \label{Xbc_FF_initial-scale_Q_rho_1_drt}
 \rho_{1,\,{\rm [direct]}}^{(Q,\tilde{Q})}(z) &\,=\, 
 -8\, {\cal R}_{\tilde{Q}/Q}^4\, (-6 + 6 z - 6 z^2 + z^3) 
 - 
 16\, {\cal R}_{\tilde{Q}/Q}^3\, (-12 + 14 z - 12 z^2 + 3 z^3) \\[0.15cm]
 &\,+\, 
 4\, {\cal R}_{\tilde{Q}/Q}^2\, (72 - 112 z + 102 z^2 - 37 z^3 + 6 z^4)  \\[0.15cm] 
 &\,+\,
 8\, {\cal R}_{\tilde{Q}/Q}\, (24 - 52 z + 54 z^2 - 22 z^3 + 3 z^4) \\[0.15cm]
 &\,+\, 48 - 144 z + 168 z^2 - 84 z^3 + 19 z^4 \;,
\end{split}
\end{equation}
\begin{equation}
\begin{split}
 \label{Xbc_FF_initial-scale_Q_rho_2_drt}
 \rho_{2,\,{\rm [direct]}}^{(Q,\tilde{Q})}(z) &\,=\, 
 6\, {\cal R}_{\tilde{Q}/Q}^2\, (2 - z + 2 z^2) 
 + 
 4\, {\cal R}_{\tilde{Q}/Q}\, (6 - 4 z + 3 z^2) 
 +
 12 - 18 z + 11 z^2 \;,
\end{split}
\end{equation}
and
\begin{equation}
\begin{split}
 \label{Xbc_FF_initial-scale_Q_rho_3_drt}
 \rho_{3,\,{\rm [direct]}}^{(Q,\tilde{Q})}(z) \,=\, 1 \;,
\end{split}
\end{equation}
where ${\cal R}_{\tilde{Q}/Q} \equiv m_{\tilde{Q}}/m_Q$\,.

Relative to Ref.~\cite{Nejad:2021mmp}, our formulation differs both in the explicit evaluation of ${\mathcal N^{(Q)}_{\rm [direct]}}$ and in the treatment adopted for the $\vqTTa$ parameter.
Within the Suzuki model, $\vqTTa$ governs the transverse-momentum profile of the bound state and effectively enters the nonperturbative wave-function component (F-NPWF).
As such, $\vqTTa$ should be regarded as a phenomenological parameter whose value must be constrained through physically motivated criteria.
As pointed out in Ref.~\cite{GomshiNobary:1994eq}, increasing $\vqTTa$ progressively shifts the FF peak toward the small-$z$ region, while simultaneously reducing the normalization of its bulk contribution.

In Ref.~\cite{Nejad:2021mmp}, the initial-scale heavy-quark FF into doubly heavy $\QXQq$ tetraquarks was derived using the choice $\vqTTa = 1\mbox{ GeV}^2$, interpreted there as an upper estimate of the average transverse momentum squared.
For the present $\QXbc$ study, we instead adopt a revised prescription for $\vqTTa$, intended as a balanced phenomenological choice compatible with the exploratory character of our analysis.
This strategy follows the tuning procedure previously introduced in studies of all-heavy $\TQQ$ tetraquark fragmentation~\cite{Celiberto:2024mab}.
There, $\vqTTa$ was adjusted in order to ensure $\langle z \rangle \gtrsim 0.4$ together with a comparable normalization between heavy-quark- and gluon-initiated channels.
Applying the same criterion to the $\QXbc$ system, we select
\begin{equation}
\label{eq:vqTTa_Xbc}
\vqTTa_{\QXbc} \equiv 70 \, \text{GeV}^2 \;.
\end{equation}

Although substantially larger than typical hadronic scales, this quantity should be interpreted as an effective parameter rather than as a genuinely soft nonperturbative scale.
In fully heavy systems, the absence of light constituents prevents the bound state from being characterized by a single soft scale directly associated with $\LQCD$, while the presence of multiple heavy quarks naturally favors a broader intrinsic momentum distribution inside the hadron.
From this perspective, $\vqTTa$ parametrizes the intrinsic momentum spread of the fully heavy configuration and should not be interpreted as a direct measure of low-scale nonperturbative dynamics.
Instead, it effectively encodes modulation effects associated with the formation of a compact fully heavy Fock state, where the relevant dynamical scale emerges from the interplay between heavy-quark masses and binding interactions, rather than from $\LQCD$ alone. 
This interpretation remains consistent with early studies of heavy-flavor fragmentation~\cite{Suzuki:1977km,Bjorken:1977md}, where the large-$z$ behavior is primarily controlled by the heavy-quark mass.
However, in fully heavy exotic systems such as $\QXbc$, the absence of light degrees of freedom prevents a purely kinematic determination of the FF peak position, thereby motivating a phenomenological treatment of the $\vqTTa$ parameter.

\subsection{Diquark fragmentation}
\label{ssec:FFs-Q-diquark}

The diquark picture is based on the idea that two quarks can form a correlated colored subsystem acting as an effective constituent inside a larger hadronic configuration~\cite{Gell-Mann:1964ewy}.
This framework has long been employed in hadron spectroscopy and heavy-flavor phenomenology, where it provides a convenient description of multiquark dynamics and baryonic structures~\cite{Maiani:2004vq,Jaffe:2003sg,Guo:2013xga}.
Depending on the spin configuration of the correlated quark pair, diquarks can be classified as scalar ($J=0$) or axial-vector ($J=1$) states.
Related spectator-type constructions are also widely used to model spin-dependent quark~\cite{Bacchetta:2008af} and gluon distributions~\cite{Bacchetta:2020vty,Bacchetta:2024fci}, where the composite nature of the effective subsystem is encoded through phenomenological form factors.

Scalar configurations are typically associated with a simpler single-form-factor structure, while axial-vector systems require a richer tensor decomposition.
Quark-diquark approaches have been extensively applied to heavy-flavor baryon fragmentation~\cite{Ma:2001ri,Yang:2002gh,Adamov:1997yk,MoosaviNejad:2017rvi,Delpasand:2019xpk}, and were later generalized to multiquark systems such as tetraquarks and pentaquarks~\cite{Faustov:2020qfm,Maiani:2015vwa}.
Within this picture, the diquark-antidiquark configuration effectively reduces the multiquark dynamics to a simpler two-body bound-state problem, thus providing a more tractable framework for the study of fragmentation mechanisms and bound-state formation.
At sufficiently large transverse momentum, fragmentation contributions are expected to dominate over recombination mechanisms, thereby motivating a ZM-VFNS description based on DGLAP-evolved FFs.

\begin{figure*}[!t]
\centering
\includegraphics[width=0.475\textwidth]{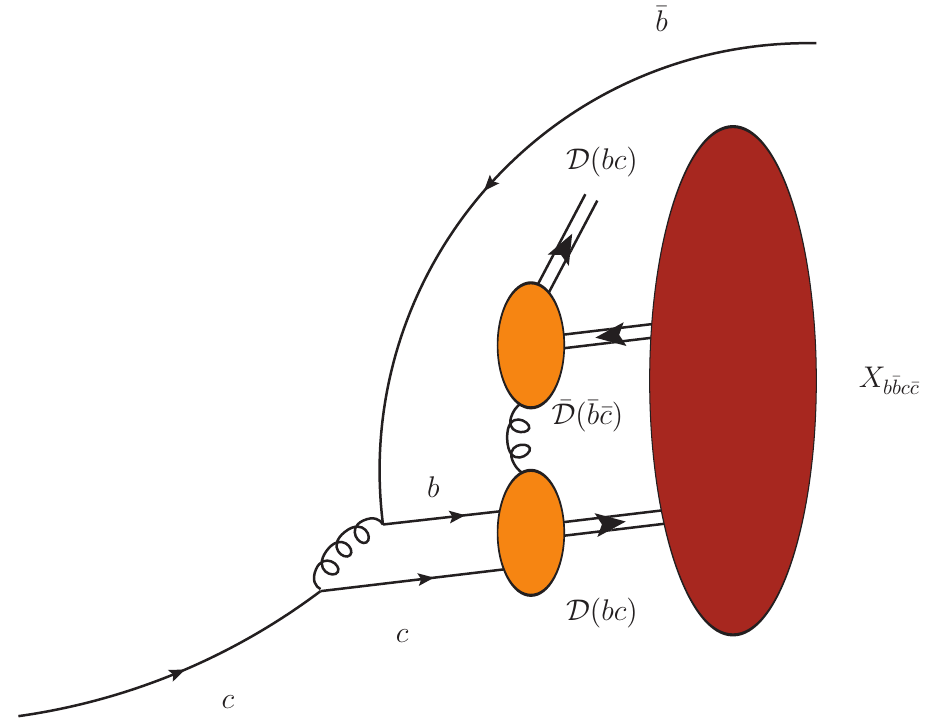}
\hspace{0.00cm}
\includegraphics[width=0.475\textwidth]{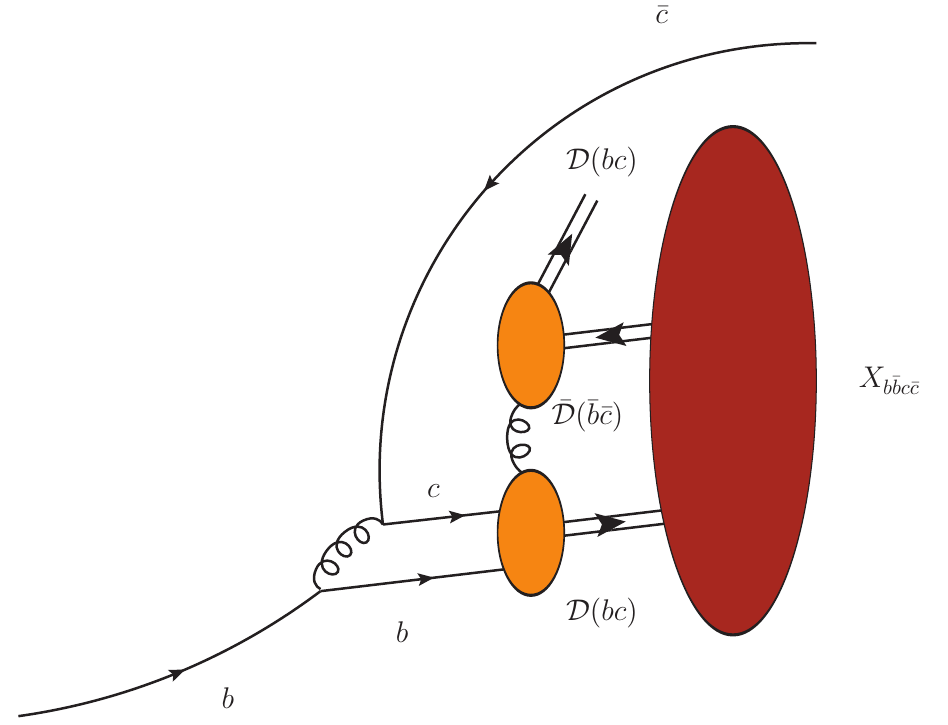}

\caption{Representative LO diagrams for the collinear fragmentation of a constituent charm (left) or bottom (right) quark into a color-singlet $\QXbc$ tetraquark in the scalar-diquark fragmentation channel.
The double fermion lines represent the tightly bound scalar diquark ${\cal D}(bc)$ and antidiquark $\bar{\cal D}(\bar b\bar c)$ configurations entering the sequential fragmentation mechanism.
The orange blobs denote the corresponding effective nonpointlike diquark vertices, while the firebrick-shaded blob represents the nonperturbative hadronization component entering the tetraquark FF.
Feynman diagrams were produced with {\tt JaxoDraw 2.0}~\cite{Binosi:2008ig}.}
\label{fig:Xbc_FF_sdq}
\end{figure*}

We now outline our strategy for modeling the initial-scale input associated with heavy-quark fragmentation into an $S$-wave color-singlet $\QXbc$ tetraquark in the scalar-diquark configuration.
As in the direct-fragmentation case, the calculation is performed within the Suzuki framework~\cite{Suzuki:1977km}, but the leading configuration is described as a compact diquark-antidiquark, schematically represented as $| {\cal D}(Q\tilde{Q}) ,\bar{\cal D}(Q\tilde{Q}) \rangle$, where ${\cal D}(Q\tilde{Q})$ denotes a colored mixed heavy diquark.
At LO, the fragmentation process proceeds through the perturbative transition
\begin{equation}
\label{eq:splitting_sdq}
[Q \,\to\, | {\cal D}(Q\tilde{Q}) \,\bar{\cal D}(\bar{Q}\bar{\tilde{Q}}) \rangle \,+\, {\cal D}(Q\tilde{Q}) \,+\, \bar{\tilde{Q}}] \;,
\end{equation}
as illustrated in Fig.~\ref{fig:Xbc_FF_sdq}.
The left and right panels correspond to the charm- and bottom-initiated fragmentation channels, respectively.

Within QCD, two quarks can combine into a color-sextet or a color-antitriplet configuration,
\begin{equation}
 \label{color_3c_x_3c_6}
 3_c \otimes 3_c = \bar{3}_c \oplus 6_c \; .
\end{equation}
Within the one-gluon-exchange picture, the interaction is attractive in the color-antitriplet channel, whereas the color-sextet configuration is disfavored~\cite{Gell-Mann:1964ewy,    Jaffe:1976ig,Maiani:2004vq}.
Accordingly, the constituent heavy diquark is assumed to transform as a color-antitriplet state, while the corresponding antidiquark transforms as a color-triplet state,
\begin{equation}
 \label{color_3c_x_3c_8}
 \bar{3}_c \otimes 3_c = 1_c \oplus 8_c \; ,
\end{equation}
so that their combination naturally contains a color-singlet component, identified with the physical tetraquark state.
Throughout this work we restrict ourselves to scalar diquarks, consistently with the assumed
$J^{PC}=0^{++}$ quantum numbers of the tetraquark states considered in this work.

The fragmentation mechanism can be conveniently interpreted as an effective two-step process.
The fragmenting heavy quark first produces an intermediate colored diquark ${\cal D}(Q\tilde{Q})$ together with a spectator antiquark $\bar{\tilde{Q}}$ through a short-distance QCD interaction.
The intermediate diquark subsequently fragments into the physical color-singlet tetraquark through the formation of a ${\cal D}(Q\tilde{Q})\bar{\cal D}(\bar{Q}\bar{\tilde{Q}})$ bound state, accompanied by an additional outgoing diquark ${\cal D}(Q\tilde{Q})$.
The antidiquark therefore enters the tetraquark bound state, whereas the additional diquark and the spectator antiquark belong to the accompanying final state.

Within the Suzuki framework~\cite{Suzuki:1977km}, we adopt the diquark-based construction proposed in Ref.~\cite{MoosaviNejad:2025leh} for the initial-scale FF.
As in the direct-fragmentation case, the relative motion of the constituents inside the tetraquark is neglected, consistently with the collinear approximation adopted throughout this work, so that the bound-state dynamics is entirely encoded in the wave function at the origin.
Here and in the following, $Q$ denotes the fragmenting heavy quark, while $\tilde{Q}$ labels the second heavy constituent entering the diquark-antidiquark bound state.
Accordingly, $(Q,\tilde{Q})=(c,b)$ for the $[c \to \QXbc]$ fragmentation channel, whereas $(Q,\tilde{Q})=(b,c)$ for the $[b \to \QXbc]$ transition.
Assuming charge-conjugation symmetry between heavy-quark and heavy-antiquark fragmentation channels, the leading-power initial-scale FF for the scalar-diquark mechanism is given by
\begin{equation}
\label{Xbc_FF_initial-scale_Q_sdq}
D_{Q,\,{\rm [diquark]}}^{\QXbc}(z,\mu_{F,0}^{(Q)}) \,=\,
\int_{z}^{1}\frac{\drv \zeta}{\zeta}\,
D_{Q}^{\mathcal D}\!\left(\frac{z}{\zeta}\right)
D_{\mathcal D}^{\QXbc}(\zeta)
\,.
\end{equation}
where $\zeta$ denotes the longitudinal momentum fraction transferred from the intermediate scalar diquark to the produced tetraquark, while $z/\zeta$ is the corresponding momentum fraction carried by the diquark with respect to the fragmenting heavy quark.

Equation~\eqref{Xbc_FF_initial-scale_Q_sdq} follows the same factorization pattern discussed in Sec.~\ref{ssec:FFs-intro} for singly heavy hadrons and quarkonia.
The perturbative function $D_{Q}^{\mathcal D}$ plays a role analogous to the SDC, describing the perturbative production of the colored heavy constituent, whereas $D_{\mathcal D}^{\QXbc}$ encodes its subsequent nonperturbative transition into the physical color-singlet tetraquark.
Their convolution therefore provides the leading-power initial condition for heavy-quark fragmentation in the scalar-diquark picture.

The physical interpretation of Eq.~\eqref{Xbc_FF_initial-scale_Q_sdq} is schematically illustrated in Fig.~\ref{fig:Xbc_FF_sdq}.
For definiteness, we first discuss the charm-initiated fragmentation channel, corresponding to $(Q,\tilde{Q})=(c,b)$ and represented by the left diagram.
The fragmenting charm quark first undergoes the perturbative transition described by $D_{Q}^{\mathcal D}$, producing the intermediate scalar diquark ${\cal D}(bc)$ together with the spectator antiquark $\bar b$.
This diquark constitutes the intermediate colored degree of freedom entering the convolution in Eq.~\eqref{Xbc_FF_initial-scale_Q_sdq} and is represented by the lower double-fermion line in Fig.~\ref{fig:Xbc_FF_sdq}, with the adjacent orange blob denoting its effective nonpointlike vertex.

The subsequent fragmentation of the intermediate diquark, encoded in $D_{\mathcal D}^{\QXbc}$, proceeds through the production of the antidiquark $\bar{\cal D}(\bar b\bar c)$, represented by the upper double-fermion line, whose corresponding orange blob denotes the associated effective diquark vertex.
The binding of ${\cal D}(bc)$ and $\bar{\cal D}(\bar b\bar c)$ finally gives rise to the physical color-singlet tetraquark.
The right diagram corresponds to the bottom-initiated fragmentation channel, namely $(Q,\tilde{Q})=(b,c)$, while preserving the same convolution structure.

The two components entering the initial-scale FF in Eq.~\eqref{Xbc_FF_initial-scale_Q_sdq} are now specified separately.
We first consider the perturbative contribution describing the production of the intermediate scalar diquark.
Building upon the Suzuki-based calculation of Ref.~\cite{MoosaviNejad:2017rvi}, we recast the perturbative component entering Eq.~\eqref{Xbc_FF_initial-scale_Q_sdq} into the following compact form:
\begin{equation}
\label{Xbc_FF_initial-scale_Q-to-D_sdq}
 D_Q^{\mathcal D}(\xi)
 =
 {\mathcal N^{(Q\to{\cal D})}_{\rm [diquark]}}\,
 \frac{\xi(1-\xi)^3}
 {\left[\Xi^{(Q,\tilde Q)}_{\rm [diquark]}(\xi)\right]^2}
 \sum_{l=0}^{2}
 \xi^{2l}\,
 \rho_{l,\,{\rm [diquark]}}^{(Q,\tilde Q)}(\xi)
 \left\{ \frac{\vqTTa}{m_Q^2} \right\}^{\!l}
\;,
\end{equation}
with
\begin{equation}
\label{Xbc_FF_initial-scale_Q-to-D_N_sdq}
{\mathcal N^{(Q\to{\cal D})}_{\rm [diquark]}}
=
\frac{2}{3}
\left[
 \pi\,
 f_{\mathcal D}\,
 C_F\,
 \alpha_s(3\mu_X+m_{\tilde Q})
\right]^2
\;.
\end{equation}
Here, $\mu_X=m_Q+m_{\tilde Q}$, as defined in Sec.~\ref{ssec:FFs-Q-direct}, while $f_{\mathcal D}$ denotes the effective nonperturbative normalization parameter controlling the formation of the tightly-bound heavy scalar diquark. 
Its physical interpretation and phenomenological determination will be discussed in the following.
The scalar-diquark mass is evaluated in the constituent approximation,
\begin{equation}
 \label{mD}
 m_{\mathcal D}\equiv m_{\bar{\mathcal D}}
 =
 m_Q+m_{\tilde Q}\;,
\end{equation}
where binding-energy effects are neglected consistently with the Suzuki prescription.
The remaining conventions, including the definition of $C_F=(N_c^2-1)/(2N_c)$ and the heavy-quark masses $m_c=1.5$~GeV and $m_b=4.9$~GeV, follow Sec.~\ref{ssec:FFs-Q-direct}.
One defines
\begin{equation}
 \label{Xbc_FF_initial-scale_Q-to-D_Xi_sdq}
 \Xi^{(Q,\tilde{Q})}_{\rm [diquark]}(\xi) \,=\,
 \frac{{\mathcal F}_{\rm [diquark]}^{(Q,\tilde{Q})}(\xi)}
 {m_Q^5 \, m_{\tilde{Q}}}
 \;,
\end{equation}
and 
\begin{equation}
\begin{split}
\label{Xbc_FF_initial-scale_Q-to-D_F_sdq}
 {\mathcal F}_{\rm [diquark]}^{(Q,\tilde{Q})}(\xi)
\,&=\,
 \left\{
 (\xi-1)\left(m_{\mathcal D}^2-\xi m_Q^2\right)
 -\xi\left(m_{\tilde Q}^2+\xi\vqTTa\right)
 \right\}
 \\[0.15cm]
 \,&\times\,
 \left\{
 \xi^2\vqTTa
 +\left[m_{\mathcal D}(\xi-1)-\xi m_{\tilde Q}\right]^2
 \right\}
\;.
\end{split}
\end{equation}
The definition of $\Xi_{\rm [diquark]}^{(Q,\tilde Q)}$ is chosen so that all coefficient functions $\rho_{l,\rm [diquark]}^{(Q,\tilde Q)}$ are dimensionless.
The expressions for the $\rho_{l,\,{\rm [diquark]}}^{(Q,\tilde{Q})}(z)$ coefficients in Eq.~\eqref{Xbc_FF_initial-scale_Q-to-D_sdq} are
\begin{equation}
\begin{split}
\label{Xbc_FF_initial-scale_Q-to-D_rho0_sdq}
 \rho_{0,\,{\rm [diquark]}}^{(Q,\tilde Q)}(\xi)
 &\,=\,
 \frac{{\cal R}_{Q/\tilde{Q}}}{m_Q^4}
 \,
 \bigg \{
 (1-\xi)^2
 m_{\mathcal D}^{4}
 -
 2\xi(1-\xi) \,
 m_{\mathcal D}^{3}
 \left[
 (1-\xi)m_Q-m_{\tilde Q}
 \right]
\\[0.10cm]
 &\,+\,
 \xi^2 \,
 m_{\mathcal D}^2
 \left[
 3(1-\xi)^2m_Q^2
 +3m_{\tilde Q}^2
 -8(1-\xi)m_Qm_{\tilde Q}
 \right]
\\[0.10cm]
 &\,+\,
 2\xi^3 \, 
 m_{\mathcal D}
 m_Q
 m_{\tilde Q}
 \left[
 (1-\xi)m_Q-m_{\tilde Q}
 \right]
 +
 \xi^4 \,
 m_Q^2m_{\tilde Q}^2
 \bigg \}
 \;,
\end{split}
\end{equation}
\begin{equation}
\begin{split}
\label{Xbc_FF_initial-scale_Q-to-D_rho1_sdq}
 \rho_{1,\,{\rm [diquark]}}^{(Q,\tilde Q)}(\xi)
 &\,=\,
 \frac{{\cal R}_{Q/\tilde{Q}}}{m_Q^2}
 \,
 \bigg \{
 m_{\mathcal D}^2
 \left(
 2+3\xi^2-2\xi
 \right)
 -
 2\xi m_{\mathcal D}
 \left[
 m_Q-(1-\xi)m_{\tilde Q}
 \right]
\\[0.10cm]
 &\,+\,
 \xi^2
 \left(
 m_Q^2+m_{\tilde Q}^2
 \right)
 \bigg \}
 \;,
\end{split}
\end{equation}
and
\begin{equation}
\label{Xbc_FF_initial-scale_Q-to-D_rho2_sdq}
 \rho_{2,\,{\rm [diquark]}}^{(Q,\tilde Q)}(\xi)
 =
 {\cal R}_{Q/\tilde{Q}}
 \;,
\end{equation}
\\[0.10cm]
where ${\cal R}_{Q/\tilde{Q}} \equiv m_Q/m_{\tilde{Q}} = \left({\cal R}_{\tilde{Q}/Q}\right)^{-1}$\,.

We now discuss the second component entering the initial-scale FF in Eq.~\eqref{Xbc_FF_initial-scale_Q_sdq}, which describes the fragmentation of the intermediate scalar diquark into the physical tetraquark.
According to Ref.~\cite{MoosaviNejad:2025leh}, the corresponding LO contribution reads
\begin{equation}
\label{Xbc_FF_initial-scale_D-to-X_sdq}
 D_{\mathcal D}^{\QXbc}(\zeta)
 =
 \frac{{\mathcal N^{({\mathcal D}\to{\QXbc})}_{\rm [diquark]}}
 \,
 \Lambda_{\cal D}^{4}\,
 m_{\QXbc}^{2}
 \,
 \,\zeta(1-\zeta)^2
 \left[\zeta^2\vqTTa
 +m_{\mathcal D}^2(\zeta+2)^2\right]}
 {\left[\zeta^2\vqTTa
 +(\zeta-2)^2m_{\mathcal D}^2\right]
 \left[\zeta^2\!\left(m_{\mathcal D}^2
 +\vqTTa\right)
 +(1-\zeta)m_{\QXbc}^2\right]^2}
 \;,
\end{equation}
where $m_{\QXbc}$ denotes the tetraquark mass.
Since $\QXbc$ is treated here as a hypothetical state, binding-energy effects are neglected, and the constituent approximation gives
\begin{equation}
\label{mX}
 m_{\QXbc}
 =
 2(m_Q+m_{\tilde Q})
 \equiv
 2m_{\mathcal D} 
 \;.
\end{equation}
The parameter $\vqTTa$ denotes the phenomenological transverse-momentum parameter entering the Suzuki prescription, as discussed in Sec.~\ref{ssec:FFs-Q-direct}.

The overall normalization factor $\mathcal N^{({\mathcal D}\to{\QXbc})}_{\rm [diquark]}$ in Eq.~\eqref{Xbc_FF_initial-scale_D-to-X_sdq}, inherited from the short-distance calculation, reads
\begin{equation}
\label{Xbc_FF_initial-scale_D-to-X_N_sdq}
 {\mathcal N^{({\mathcal D}\to{\QXbc})}_{\rm [diquark]}}
 =
 \left[
 \frac{\pi}{4\sqrt{2}}\,
 \alpha_s(m_{\QXbc})\,
 C_F\,
 m_{\mathcal D}\,
 f^{(X)}_{\rm [diquark]}
 \right]^2
 \;,
\end{equation}
where $f^{(X)}_{\rm [diquark]}$ denotes the effective nonperturbative normalization parameter modulating the tetraquark bound-state wave function within the sequential scalar-diquark fragmentation mechanism, while $\Lambda_{\cal D}$ is the effective scale parameter entering the scalar-diquark form factor.

Relative to the scalar-diquark construction of Ref.~\cite{MoosaviNejad:2025leh}, our formulation differs in the explicit reorganization of the two components entering the convolution in Eq.~\eqref{Xbc_FF_initial-scale_Q_sdq}, as well as in the prescriptions adopted for their normalization parameters and for $\vqTTa$.
For the latter, we follow the same phenomenological strategy introduced in the direct-fragmentation channel and use the common value $\vqTTa_{\QXbc}=70\,\text{GeV}^{2}$ defined in Eq.~\eqref{eq:vqTTa_Xbc}.
This prescription provides a uniform treatment of intrinsic transverse-momentum dependence across the two fragmentation mechanisms and enables a direct comparison between their corresponding initial-scale FFs.

The normalization factors entering the initial-scale FFs have also been reformulated with respect to the original Suzuki-based implementations.
As already anticipated for the direct-fragmentation channel, we evaluate explicitly the normalization factors appearing in
Eqs.~\eqref{Xbc_FF_initial-scale_Q_N_drt},
\eqref{Xbc_FF_initial-scale_Q-to-D_N_sdq},
and
\eqref{Xbc_FF_initial-scale_D-to-X_N_sdq},
thus making their dependence on the underlying phenomenological inputs fully explicit.
These normalization factors involve the effective nonperturbative normalization parameters
$f^{(X)}_{\rm [direct]}$,
$f_{\mathcal D}$,
and
$f^{(X)}_{\rm [diquark]}$,
together with the effective scalar-diquark scale parameter
$\Lambda_{\cal D}$.
We now turn to the physical interpretation, dimensional conventions, and phenomenological determination of these effective parameters.

\begin{table*}[!t]
\centering
\small
\setlength{\tabcolsep}{8pt}
\renewcommand{\arraystretch}{1.40}
\begin{tabular}{
>{\raggedright\arraybackslash}m{2.9cm}
>{\justifying\arraybackslash\setlength{\parindent}{0pt}}p{5.5cm}
>{\justifying\arraybackslash\setlength{\parindent}{0pt}}p{6.5cm}
}
\toprule
\textbf{Parameter [units]} &
\textbf{Role in the FF} &
\textbf{Prescription / Values} \\
\midrule

$f_{\mathcal M}$ [GeV] &
Heavy-light meson decay constant used exclusively to construct
$f_{\mathcal D}$. &
Phenomenological input.
See text. \\[0.25cm]
\addlinespace[0.15cm]

$f_{\mathcal D}$ [GeV$^2$] &
Effective diquark decay constant entering the normalization of the perturbative
$[Q \to \mathcal D]$
FF. &
Constructed according to
$f_{\mathcal D}=m_{\mathcal D}f_{\mathcal M}/(2\sqrt2)$. \\[0.25cm]
\addlinespace[0.15cm]

$f^{(X)}_{\rm [direct]}$ [1] &
Effective normalization parameter entering the direct formation of the tetraquark in the
$[Q \to X]$
fragmentation channel. &
Assumed proportional to the diquark normalization parameter,
$f^{(X)}_{\rm [direct]}=\kappa_Qf_{\mathcal D}$,
with $\kappa_Q\simeq1/{\rm GeV}^2$. \\[0.25cm]
\addlinespace[0.15cm]

$f^{(X)}_{\rm [diquark]}$ [GeV$^{-2}$] &
Effective normalization parameter entering the nonperturbative transition
$[\mathcal D \to X]$
in the scalar-diquark mechanism. &
Nonstrange values from QCD sum rules:
$f_{cu}^{(X)}=0.62\times10^{-2}$,
$f_{bu}^{(X)}=1.38\times10^{-2}$.
Strange states are obtained through
$\left[f_{{\mathcal M}(Qs)}/f_{{\mathcal M}(Qu)}\right]^2$,
while the mixed-$bc$ state is estimated through
$\sqrt{f_{cu}^{(X)}f_{bu}^{(X)}}$. \\[0.25cm]
\addlinespace[0.15cm]

$\Lambda_{\cal D}$ [GeV] &
Effective scalar-diquark form-factor scale entering the
$[\mathcal D \to X]$
fragmentation component. &
Common benchmark value
$\Lambda_{\cal D}=15~\mathrm{GeV}$
adopted for charm- and bottom-initiated fragmentation.
See text. \\

\bottomrule
\end{tabular}

\caption{Summary of phenomenological parameters entering the normalization of the initial-scale FFs in the direct and scalar-diquark fragmentation mechanisms.}
\label{tab:Xbc_FF_hadronic_parameters}
\end{table*}

\subsection{Physical motivation for  the phenomenological parameters}

The phenomenological parameters entering the normalization factors needed to describe the FFs introduced above are summarized in Table~\ref{tab:Xbc_FF_hadronic_parameters}. 
In the following, we discuss the physical motivation and the prescriptions adopted for each of them.

The heavy-light meson decay constants $f_{\mathcal M}$ provide one class of external phenomenological inputs entering our construction.
Throughout this work, we employ the representative values
$f_B=0.20$,
$f_D=0.21$,
$f_{B_s}=0.24$,
$f_{D_s}=0.25$,
and
$f_{B_c}=0.50~\mathrm{GeV}$,
consistently with modern lattice-QCD and QCD sum-rule determinations~\cite{Gelhausen:2013wia,Colquhoun:2015oha,Baker:2013mwa,FlavourLatticeAveragingGroupFLAG:2024oxs}. These quantities do not enter the FFs directly. 
Instead, they provide the phenomenological input required to determine the effective scalar-diquark decay constant governing the perturbative production of the intermediate scalar diquark.

Following the prescription adopted in Ref.~\cite{MoosaviNejad:2025leh}, we define
\begin{equation}
\label{eq:fD_prescription}
f_{\mathcal D}
=
\frac{m_{\mathcal D}\,f_{\mathcal M}}
{2\sqrt2}
\;,
\end{equation}
which follows the phenomenological assumption adopted in Ref.~\cite{MoosaviNejad:2025leh}, according to which the squared wave function at the origin of a heavy scalar diquark is related to that of the corresponding heavy quarkonium by
$|\Psi_{QQ}(0)|^2\simeq|\Psi_{Q\bar Q}(0)|^2/8$.
Within our phenomenological prescription, the factor $m_{\mathcal D}$ is retained explicitly in the definition of $f_{\mathcal D}$ to specify the adopted normalization convention.
This prescription provides a simple phenomenological connection between the nonperturbative scales governing heavy-meson and heavy-diquark formation.

The effective normalization parameter associated with the subsequent nonperturbative transition
$\mathcal D\to X$,
denoted by
$f_{\rm [diquark]}^{(X)}$,
is treated separately.
The nonstrange values
\begin{equation}
\label{eq:fX_diquark_nonstrange}
f_{{\rm [diquark]},cu}^{(X)}
=
0.62\times10^{-2},
\qquad
f_{{\rm [diquark]},bu}^{(X)}
=
1.38\times10^{-2}
\;,
\end{equation}
are taken from the available QCD sum-rule determinations for doubly heavy tetraquarks~\cite{Agaev:2018vag,Agaev:2018khe}. 
In our fragmentation framework, these numerical inputs are adopted as effective phenomenological normalization parameters
with mass dimension $-2$.
Since no corresponding determinations are presently available for the strange and mixed-$bc$ configurations, we extend these QCD sum-rule results through simple phenomenological prescriptions.

For strange systems, we assume that the dominant $SU(3)$-flavor breaking effects can be effectively absorbed into the corresponding heavy-light meson decay constants. 
Accordingly, the nonperturbative normalization is rescaled according to
\begin{equation}
\label{eq:fX_diquark_strange}
f_{{\rm [diquark]},Qs}^{(X)}
=
f_{{\rm [diquark]},Qu}^{(X)}
\left[
\frac{f_{{\mathcal M}(Qs)}}
     {f_{{\mathcal M}(Qu)}}
\right]^2
\;,
\end{equation}
which preserves the normalization of the nonstrange QCD sum-rule result while consistently accounting for the enhancement induced by replacing a light quark with a strange one through well-established hadronic inputs.

For the mixed-$bc$ configuration, no phenomenological or QCD sum-rule determination is presently available. 
We therefore adopt the prescription
\begin{equation}
\label{eq:fX_diquark_bc}
f_{{\rm [diquark]},bc}^{(X)}
=
\sqrt{
f_{{\rm [diquark]},cu}^{(X)}
\times
f_{{\rm [diquark]},bu}^{(X)}
}
\;,
\end{equation}
namely the geometric mean of the charm- and bottom-sector normalization parameters. 
This choice naturally interpolates between the two heavy-flavor sectors, is symmetric under the exchange
$b\leftrightarrow c$,
and preserves the multiplicative character of the underlying nonperturbative normalization, yielding a reference normalization naturally lying between the corresponding hidden-charm and hidden-bottom limits.

Finally, the effective normalization parameter entering the direct fragmentation mechanism is not introduced as an additional independent phenomenological input. 
Instead, we assume
\begin{equation}
\label{eq:fX_direct_prescription}
f_{\rm [direct]}^{(X)}
=
\kappa_Q\,f_{\mathcal D},
\qquad
\kappa_Q\simeq1/{\rm GeV}^2
\;,
\end{equation}
thereby relating the nonperturbative normalization of the direct fragmentation mechanism to the same hadronic scale governing scalar-diquark formation. This prescription replaces the phenomenological normalization adopted in previous Suzuki-inspired implementations by a construction directly linked to the scalar-diquark sector. 
As a result, the direct and sequential fragmentation mechanisms are placed on the same nonperturbative footing, while simultaneously reducing the number of independent phenomenological inputs entering the model.

The remaining phenomenological parameter entering the scalar-diquark fragmentation mechanism is the effective form-factor scale $\Lambda_{\mathcal D}$, which parametrizes the finite-size effects associated with the intermediate scalar diquark. In the notation of Ref.~\cite{MoosaviNejad:2025leh}, this parameter corresponds to $Q_S$ entering the scalar-diquark form factor. 
Since the scalar-diquark contribution to the FF scales as $\Lambda_{\mathcal D}^4$, the overall normalization of the sequential fragmentation mechanism exhibits a pronounced sensitivity to its numerical value.

In Ref.~\cite{MoosaviNejad:2025leh}, benchmark values of $Q_S=5$ and $7~\mathrm{GeV}$ are adopted for charm- and bottom-containing systems, respectively.
Although these values provide a useful phenomenological reference, no analogous determination is presently available for the mixed-flavor scalar diquark ${\cal D}(bc)$ relevant to the present study.
Since the same intermediate bound state ${\cal D}(bc)$ is produced in both charm- and bottom-initiated fragmentation channels, we naturally employ a common reference value,
\begin{equation}
\label{eq:LambdaD_prescription}
\Lambda_{\mathcal D}=15~\mathrm{GeV}\;,
\end{equation}
throughout this work.

This choice is motivated by exploratory phenomenological studies performed within the present framework.
We find that the benchmark values adopted in the literature lead to an almost negligible scalar-diquark contribution, particularly in the charm-initiated channel.
The reference value adopted here enhances the phenomenological visibility of the sequential mechanism while preserving the pronounced hierarchy with respect to the direct-multiquark mode.

Given the presently limited information on $\Lambda_{\mathcal D}$, we further consider the representative interval
\begin{equation}
\label{eq:LambdaD_variation}
10~\mathrm{GeV}
\leq
\Lambda_{\mathcal D}
\leq
20~\mathrm{GeV}\;.
\end{equation}
Since the scalar-diquark FF scales as $\Lambda_{\mathcal D}^4$, this range corresponds to an overall rescaling by approximately $0.20$ to $3.16$ with respect to the reference prediction.
Such variations affect only the normalization of the scalar-diquark mode, leaving its characteristic $z$ dependence unchanged and, owing to the linearity of DGLAP evolution, propagating multiplicatively to all evolution-generated partonic components.
Within this range, however, the direct-multiquark contribution remains dominant for both charm- and bottom-initiated fragmentation, so that the qualitative hierarchy between the two mechanisms is preserved.

\subsection{The {\tt TQ4Qhf1.0} functions}
\label{ssec:FFs-TQ4Qhf10}

The construction of the {\tt TQ4Qhf1.0} FFs is completed by consistently evolving the initial-scale inputs through DGLAP dynamics.
This step connects the hadron-structure-inspired modeling developed in the previous sections with the perturbative description of fragmentation at higher energy scales.

The initial scales entering the evolution are determined by the kinematics of the corresponding fragmentation mechanisms.
In the direct mode, the minimal invariant mass associated with the
$[Q \to | Q \tilde{Q} \bar{\tilde{Q}} \bar{Q} \rangle + Q]$
splitting is (see Fig.~\ref{fig:Xbc_FF_drt})
\begin{equation}
\label{eq:muF0_direct}
\mu_{F,0}^{\rm [direct]}
=
3m_Q+2m_{\tilde Q}
\;,
\end{equation}
whereas the corresponding threshold for the
$[Q \to | {\cal D}(Q\tilde{Q}) \bar{\cal D}(\bar{Q}\bar{\tilde{Q}}) \rangle + {\cal D}(Q\tilde{Q}) + Q]$
splitting in the scalar-diquark mode reads (see Fig.~\ref{fig:Xbc_FF_sdq})
\begin{equation}
\label{eq:muF0_diquark}
\mu_{F,0}^{\rm [diquark]}
=
3m_Q+4m_{\tilde Q}
\;.
\end{equation}
These scales define the natural starting points for the perturbative evolution of the two fragmentation components.

The evolution is performed within the {\HFNRevo} scheme~\cite{Celiberto:2025euy,Celiberto:2024mex}.
Originally developed in the context of the {\tt NRFF1.0} framework~\cite{Celiberto:2025euy} to describe the DGLAP evolution of collinear FFs derived from consistent NRQCD inputs for pseudoscalar quarkonia, {\HFNRevo} was subsequently generalized to accommodate a broader class of hadron-structure-inspired initial-scale models.
In its current formulation, the framework is organized around three complementary building blocks: {\it interpretation}, {\it evolution}, and {\it uncertainty quantification}. Such a scheme has already been successfully employed in several studies of exotic-hadron fragmentation, including fully heavy tetraquarks~\cite{Celiberto:2024mab,Celiberto:2024beg,Celiberto:2025dfe,Celiberto:2025ziy,Celiberto:2026kks}, hidden-heavy tetraquarks~\cite{Celiberto:2023rzw,Celiberto:2024beg}, all-heavy pentaquarks~\cite{Celiberto:2025ipt,Celiberto:2026rzi}, and triply heavy baryons~\cite{Celiberto:2025ogy,Celiberto:2026qiz}.
The present {\tt TQ4Qhf1.0} construction extends this program to all-heavy mixed-flavor tetra\-quarks.

The evolution is carried out in a $n_f=5$ realization of the ZM-VFNS, where all active partonic species, including charm and bottom, are treated as massless degrees of freedom during DGLAP evolution.
Heavy-quark masses enter the construction through the initial fragmentation conditions and the corresponding natural starting scales, while mass-dependent effects beyond those encoded in the initial inputs are not retained in the evolution kernels.
This approximation is standard in heavy-hadron fragmentation phenomenology (see, {\it e.g.}, Ref.~\cite{Cacciari:2024kaa}) and becomes increasingly reliable in the large-transverse-momentum regime, where finite-mass corrections are expected to be power suppressed.
The kinematic selections adopted in the phenomenological analysis are chosen accordingly, so as to probe a semihard region where the leading-power fragmentation picture and the ZM-VFNS treatment underlying the {\tt TQ4Qhf1.0} determination can be consistently employed (see Sec.~\ref{sec:phenomenology}).

The evolution module itself consists of two successive stages.
A semianalytic {\it expanded and decoupled evolution} ({\tt EDevo}) provides a consistent treatment of threshold effects across different partonic channels.
This is followed by the fully numerical {\it all-order evolution} ({\tt AOevo}), which performs the complete DGLAP evolution above the corresponding matching scales.

For the present {\tt TQ4Qhf1.0} construction, however, the only nonvanishing inputs at the starting scale are the constituent charm- and bottom-quark FFs.
No additional threshold activation or channel matching is therefore required before the numerical evolution.
Accordingly, the {\tt EDevo} stage is bypassed and the two heavy-quark inputs are evolved directly through {\tt AOevo}, while the remaining partonic channels are generated dynamically by DGLAP evolution.


\begin{figure*}[!t]
\centering

\includegraphics[scale=0.410,clip]{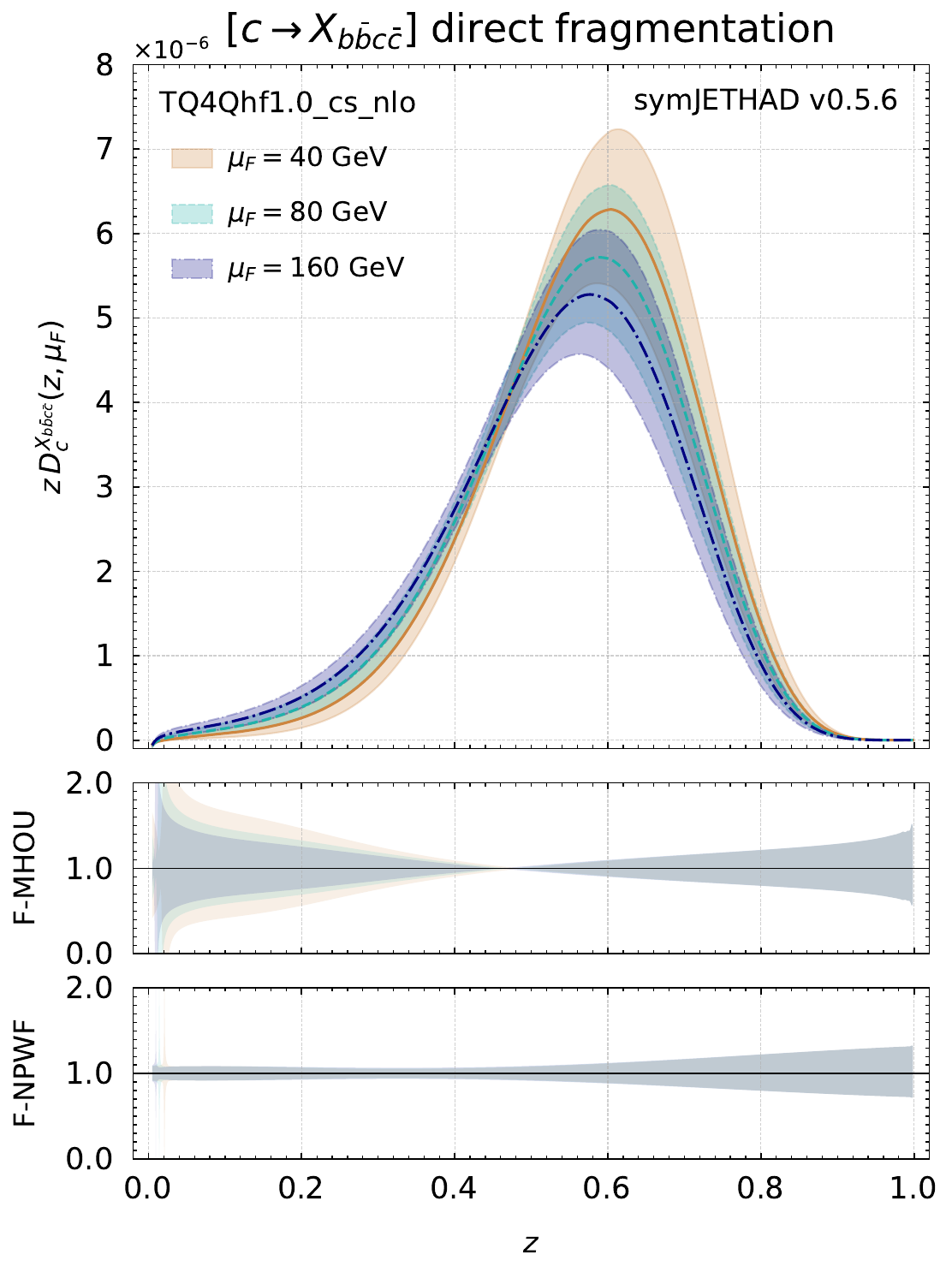}
\hspace{0.30cm}
\includegraphics[scale=0.410,clip]{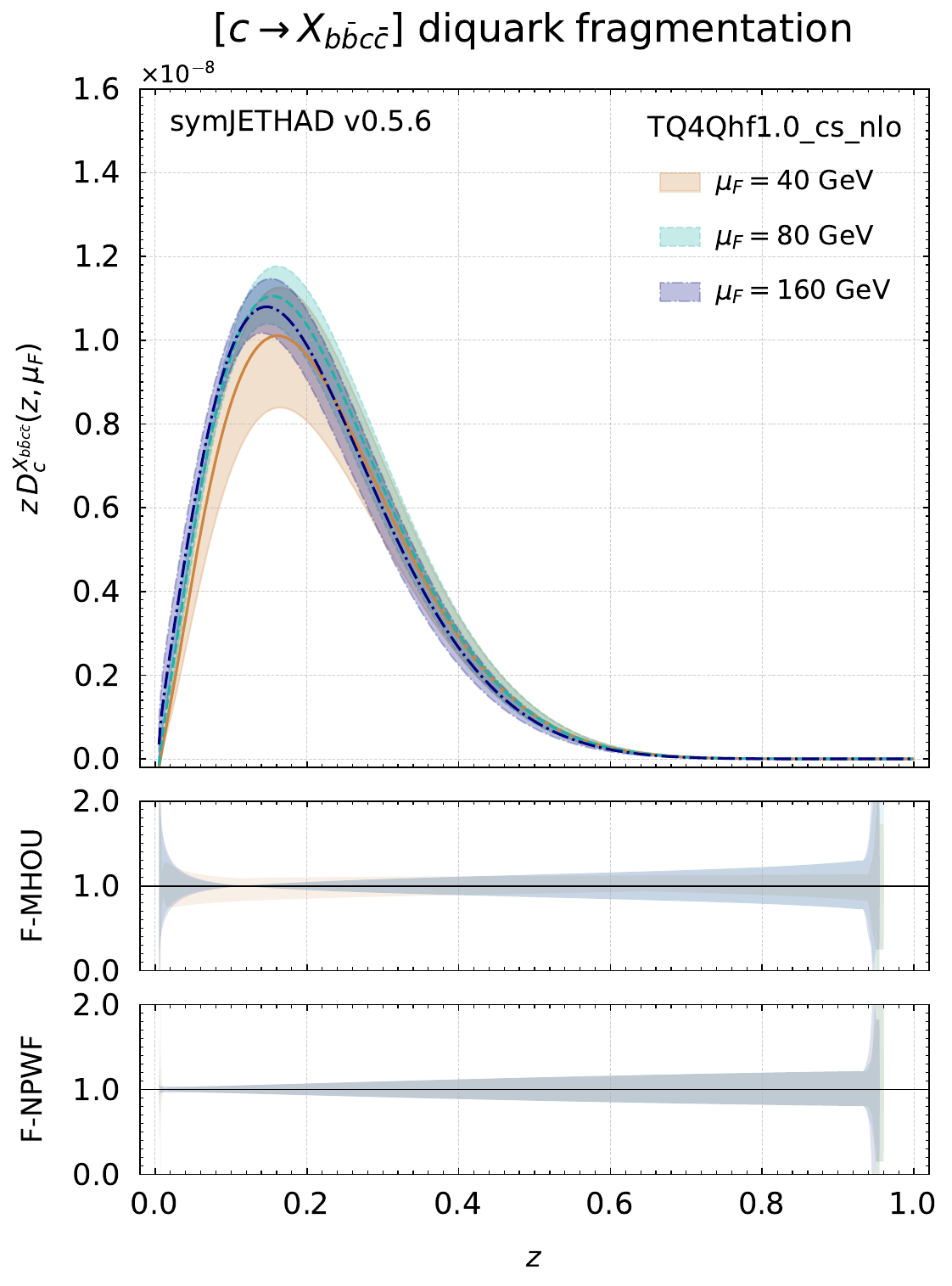}

\caption{Charm-initiated $\QXbc$ FFs in the direct-multiquark mode (left) and scalar-diquark mode (right), shown as functions of $z$.
Results are presented for $\mu_F=40$, $80$, and $160~\mathrm{GeV}$.
Solid curves denote the central predictions, while the shaded bands represent the corresponding uncertainty estimates.
The lower panels display separately the F-MHOU and F-NPWF uncertainty contributions, normalized to the central predictions.}

\label{fig:TQ4Qhf10-FFs-c}
\end{figure*}

\begin{figure*}[!t]
\centering

\includegraphics[scale=0.410,clip]{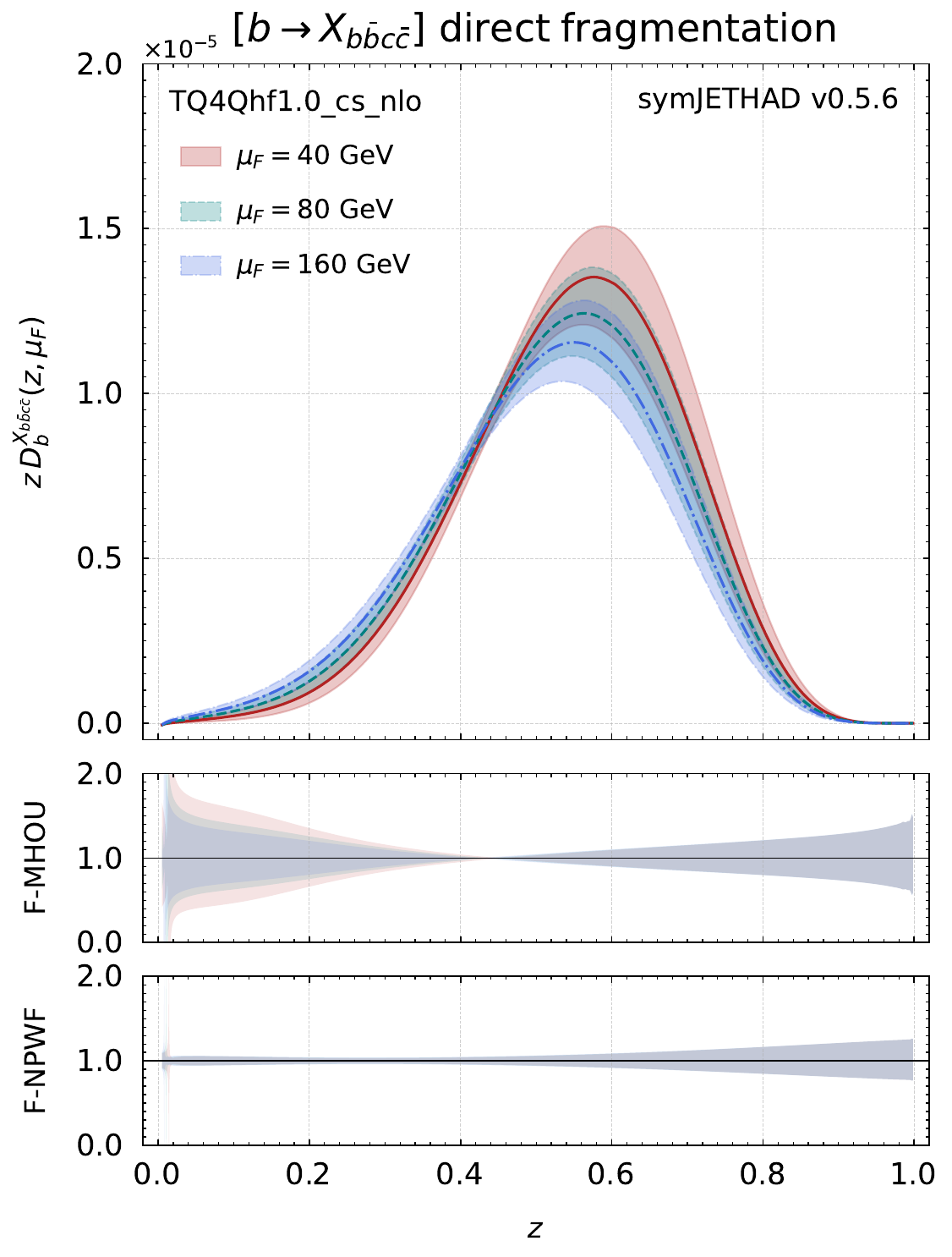}
\hspace{0.30cm}
\includegraphics[scale=0.410,clip]{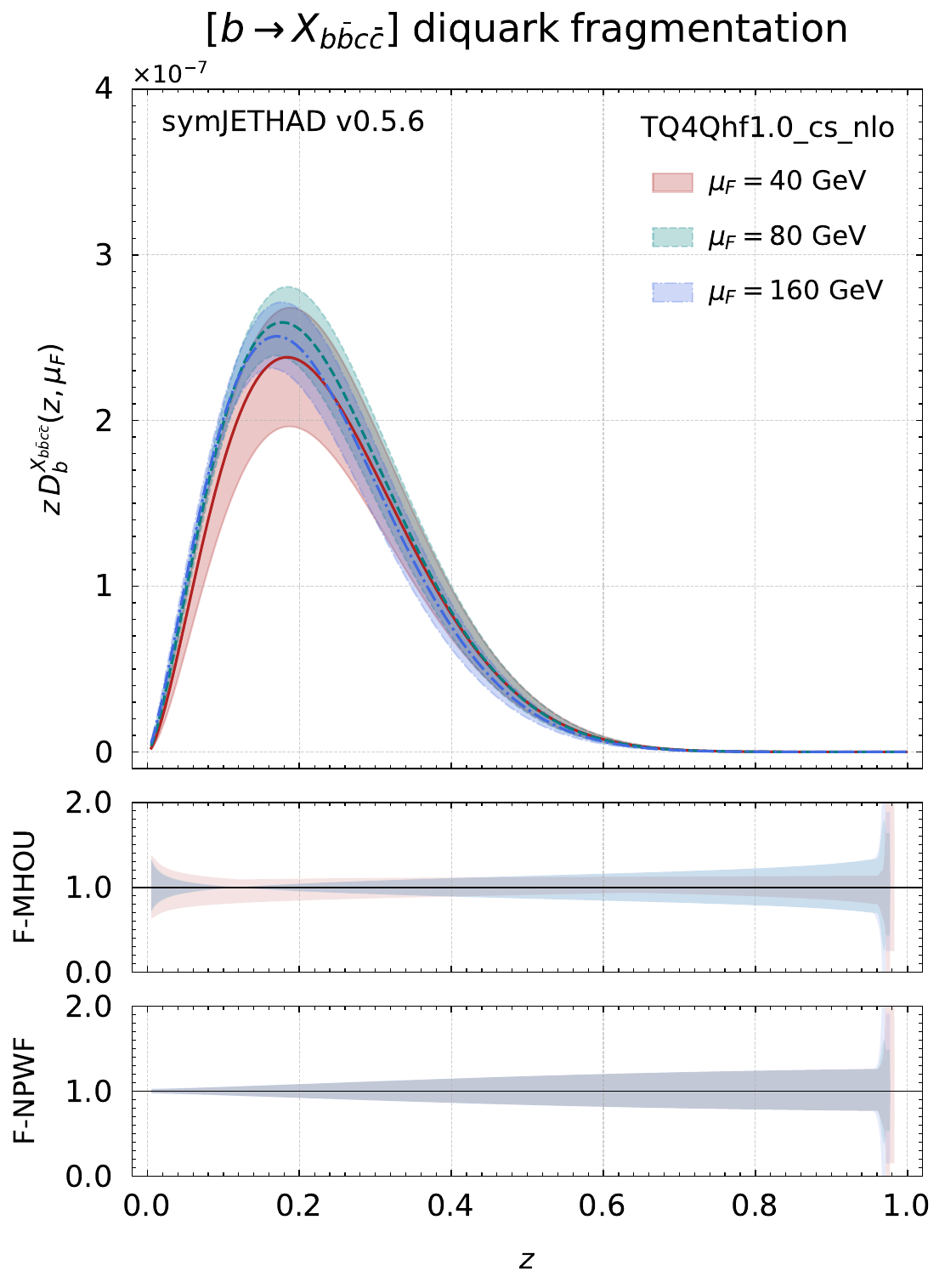}

\caption{Bottom-initiated $\QXbc$ FFs in the direct-multiquark mode (left) and scalar-diquark mode (right), shown as functions of $z$.
Results are presented for $\mu_F=40$, $80$, and $160~\mathrm{GeV}$.
Solid curves denote the central predictions, while the shaded bands represent the corresponding uncertainty estimates.
The lower panels display separately the F-MHOU and F-NPWF uncertainty contributions, normalized to the central predictions.}

\label{fig:TQ4Qhf10-FFs-b}
\end{figure*}

\begin{figure*}[!t]
\centering

\includegraphics[scale=0.410,clip]{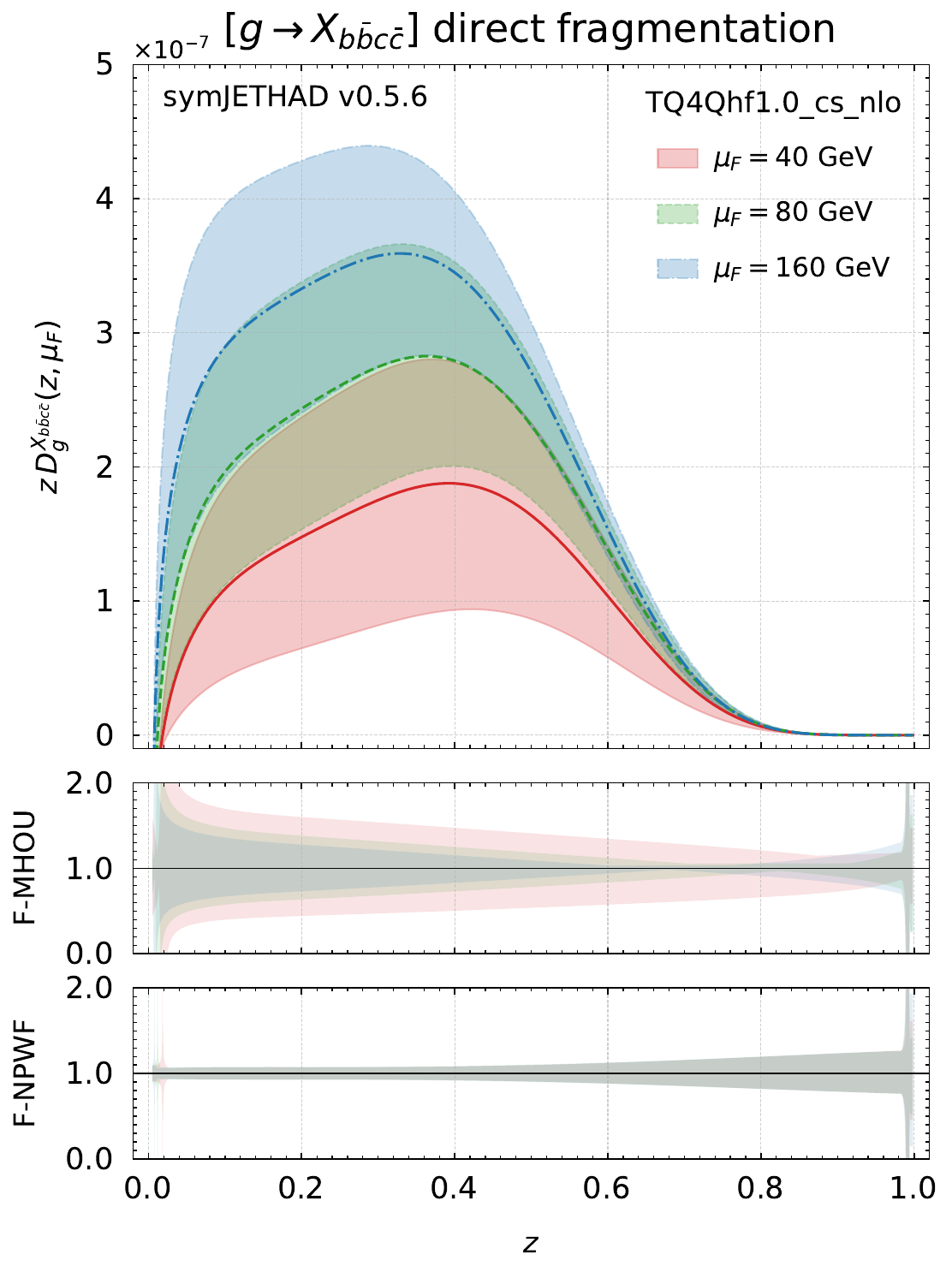}
\hspace{0.30cm}
\includegraphics[scale=0.410,clip]{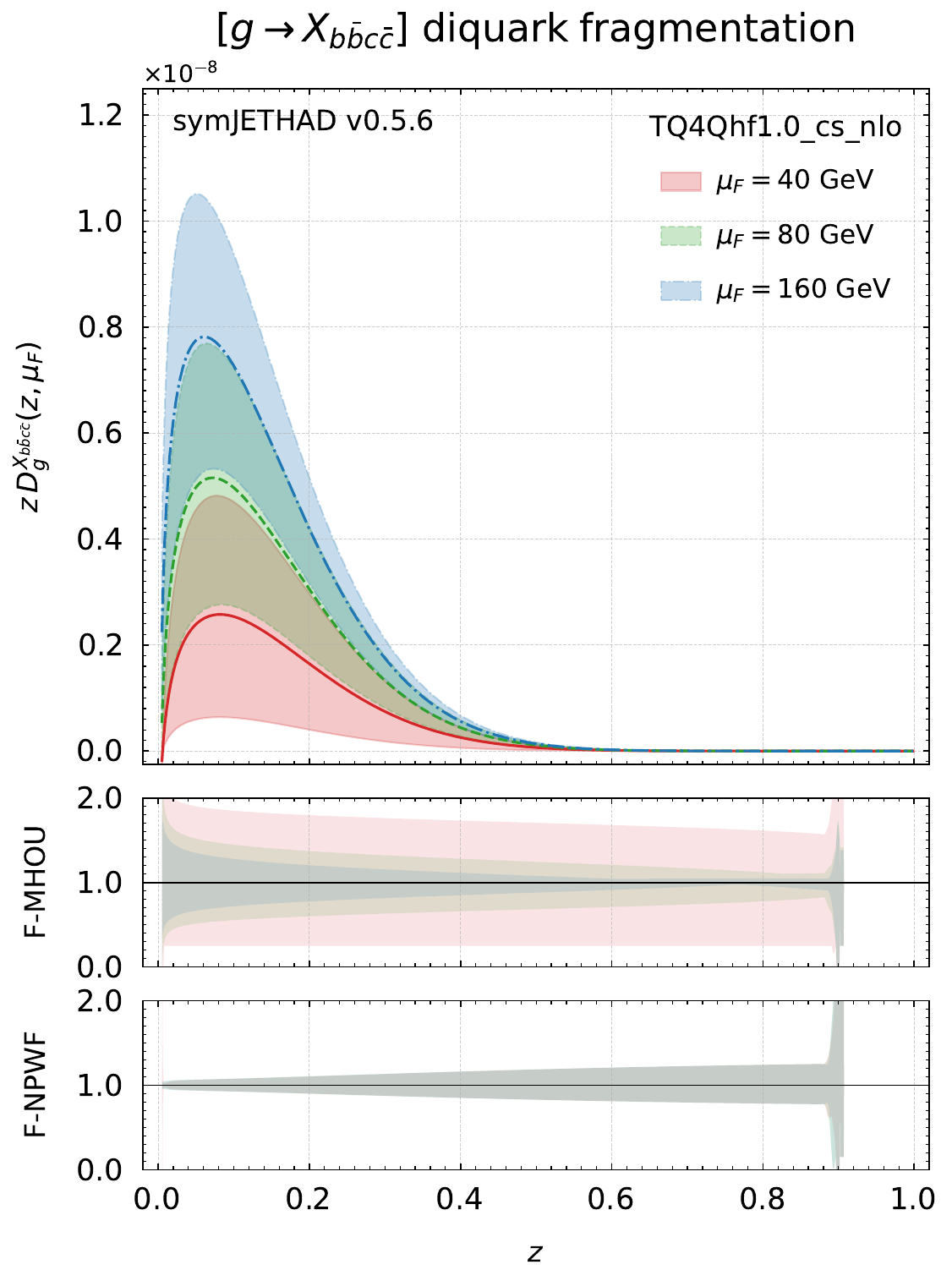}

\caption{Gluon-initiated $\QXbc$ FFs in the direct-multiquark mode (left) and scalar-diquark mode (right), shown as functions of $z$.
Results are presented for $\mu_F=40$, $80$, and $160~\mathrm{GeV}$.
Solid curves denote the central predictions, while the shaded bands represent the corresponding uncertainty estimates.
The lower panels display separately the F-MHOU and F-NPWF uncertainty contributions, normalized to the central predictions.}

\label{fig:TQ4Qhf10-FFs-g}
\end{figure*}

A limitation of the present construction is that explicit initial-scale inputs are currently available only for constituent heavy quarks.
Gluon- and light-quark-initiated FFs are therefore generated dynamically through DGLAP evolution.
Existing studies indicate that these channels are strongly suppressed at the initial scale for analogous exotic systems, such as doubly heavy tetraquarks~\cite{Nejad:2021mmp} and heavy quarkonia~\cite{Celiberto:2022dyf}.
Their explicit inclusion is left to future developments of the framework.

A distinctive feature of the {\tt TQ4Qhf1.0} set is the explicit construction of a replica ensemble encoding the dominant perturbative and nonperturbative theoretical uncertainties directly at the fragmentation-function level.
This allows the corresponding uncertainty bands to be propagated consistently throughout the entire phenomenological analysis. 
FF-intrinsic perturbative effects are estimated through missing higher-order uncertainties (F-MHOUs) by varying the evolution-ready matching scales introduced in Eqs.~(\ref{eq:muF0_direct}) and~(\ref{eq:muF0_diquark}).
Specifically, the direct- and scalar-diquark thresholds are independently rescaled by a multiplicative factor $\kappa_\mu\in[1/2,2]$, thereby probing the sensitivity of the evolved FFs to missing higher-order contributions.

Nonperturbative uncertainties are instead estimated through F-NPWF variations of the transverse-momentum parameter $\vqTTa$, which controls the intrinsic momentum distribution entering the Suzuki-inspired initial-scale construction.
The combined variation of perturbative and nonperturbative inputs defines a discrete replica ensemble, where each member corresponds to a specific choice of $\vqTTa$ and evolution-scale configuration.

These replicas should not be interpreted as a Monte Carlo statistical ensemble, but rather as a structured scan over physically motivated model variations.
Their purpose is to provide a transparent and systematically improvable estimate of the dominant theoretical uncertainties affecting the fragmentation process, while preserving the practical advantages of replica-based uncertainty propagation.
The complete correspondence between replica labels and parameter configurations is reported in Appendix~\hyperlink{app:A}{A}.
The resulting replica ensemble defines the {\tt TQ4Qhf1.0} FF set, with the central members providing the reference predictions and the remaining configurations encoding the F-MHOU and F-NPWF variations described above.

We now turn to the phenomenological structure of the evolved FFs.
Figures~\ref{fig:TQ4Qhf10-FFs-c} and~\ref{fig:TQ4Qhf10-FFs-b} display the charm- and bottom-initiated components, respectively, in the direct-multiquark and scalar-diquark modes at three representative factorization scales, $\mu_F=40$, $80$, and $160~\mathrm{GeV}$.
The side-by-side arrangement permits a direct comparison between the two fragmentation mechanisms at fixed initiating parton, while simultaneously exposing their scale dependence and the relative impact of F-MHOU and F-NPWF variations.

The corresponding gluon components are shown in Fig.~\ref{fig:TQ4Qhf10-FFs-g}.
Unlike the constituent heavy-quark FFs, the gluon channel is not independently modeled at the initial scale, but is generated dynamically through DGLAP evolution.
It therefore provides a particularly transparent probe of how the distinct direct-multiquark and scalar-diquark initial conditions propagate across the coupled partonic system.

Figures~\ref{fig:TQ4Qhf10-FFs-c} and~\ref{fig:TQ4Qhf10-FFs-b} reveal a clear separation between the direct-multiquark and scalar-diquark fragmentation patterns.
For both charm- and bottom-initiated channels, the direct-multiquark mode exhibits a comparatively hard longitudinal-momentum distribution, with a broad maximum around $z\simeq0.6$.
By contrast, the scalar-diquark mode is substantially softer and peaks at much lower momentum fractions, around $z\simeq0.15$--$0.2$.
This behavior follows naturally from the sequential structure of the scalar-diquark mechanism.
In that case, the longitudinal momentum is redistributed through the successive $[Q\to{\cal D}]$ and $[{\cal D}\to X]$ fragmentation steps, whose convolution shifts the resulting heavy-quark-to-tetraquark FF toward smaller values of $z$ with respect to the single-step direct-multiquark mechanism.
The two modes therefore differ not only in their overall normalization, but also in the characteristic longitudinal-momentum profile they imprint on the produced tetraquark.

A pronounced hierarchy in normalization is also observed when comparing the two mechanisms in Figs.~\ref{fig:TQ4Qhf10-FFs-c} and~\ref{fig:TQ4Qhf10-FFs-b}.
The separation is particularly strong in the charm-initiated channel, whereas it becomes less pronounced for bottom-initiated fragmentation.
Although the scalar-diquark contribution remains subleading in both cases, its relative weight is therefore enhanced when moving from the charm to the bottom sector.
This pattern reflects the different heavy-flavor dependence encoded in the two initial-scale constructions and survives the subsequent DGLAP evolution.

The scale dependence of the constituent heavy-quark FFs is smooth and stable over the explored range.
As $\mu_F$ increases from $40$ to $160~\mathrm{GeV}$, the direct-multiquark distributions in Figs.~\ref{fig:TQ4Qhf10-FFs-c} and~\ref{fig:TQ4Qhf10-FFs-b} undergo the expected gradual redistribution of longitudinal momentum toward lower values of $z$, accompanied by a moderate reduction of their maxima.
A milder scale dependence is observed in the scalar-diquark mode, whose characteristic low-$z$ profile remains essentially stable under evolution.

The gluon FFs shown in Fig.~\ref{fig:TQ4Qhf10-FFs-g} provide a complementary perspective on the same dynamics.
Since no independent gluon input is introduced at the starting scale, these components are generated entirely through DGLAP evolution from the constituent heavy-quark FFs.
Their magnitude consequently increases with $\mu_F$, while their longitudinal-momentum distributions are shifted toward smaller values of $z$ with respect to the corresponding heavy-quark channels.
The direct-multiquark gluon FF develops a broad maximum at intermediate $z$, whereas the scalar-diquark component is concentrated much more strongly toward the small-$z$ region.
The evolution-generated gluon sector therefore preserves, and in part amplifies, the qualitative distinction between the two fragmentation mechanisms.

The uncertainty pattern is overall well controlled across Figs.~\ref{fig:TQ4Qhf10-FFs-c}--\ref{fig:TQ4Qhf10-FFs-g}.
For the constituent heavy-quark FFs, F-MHOU effects remain moderate around the peak region and increase toward the kinematic endpoints, while F-NPWF variations produce smoother and generally smaller deformations.
A larger perturbative sensitivity is observed in the gluon channel, consistently with its purely radiative origin.
Nevertheless, the characteristic hierarchy and the distinct longitudinal-momentum profiles of the two fragmentation modes remain stable throughout the explored factorization-scale range. 
These results show that the multimodal character of the {\tt TQ4Qhf1.0} construction is preserved by perturbative evolution.
The direct-multiquark and scalar-diquark mechanisms retain clearly distinguishable signatures in the charm-, bottom-, and dynamically generated gluon components.

\section{Phenomenology at next-generation hadron colliders}
\label{sec:phenomenology}

The {\tt TQ4Qhf1.0} fragmentation set is now embedded into high-energy QCD phenomenology to assess how the direct-multiquark and scalar-diquark formation mechanisms propagate from the FF level to measurable observables.
Our analysis focuses on the production of the all-heavy mixed-flavor $\QXbc$ tetraquark, providing differential predictions, quantifying the main theoretical uncertainties, and examining the phenomenological separation between the two fragmentation modes.

\begin{figure*}[!t]
\centering
\includegraphics[width=0.65\textwidth]{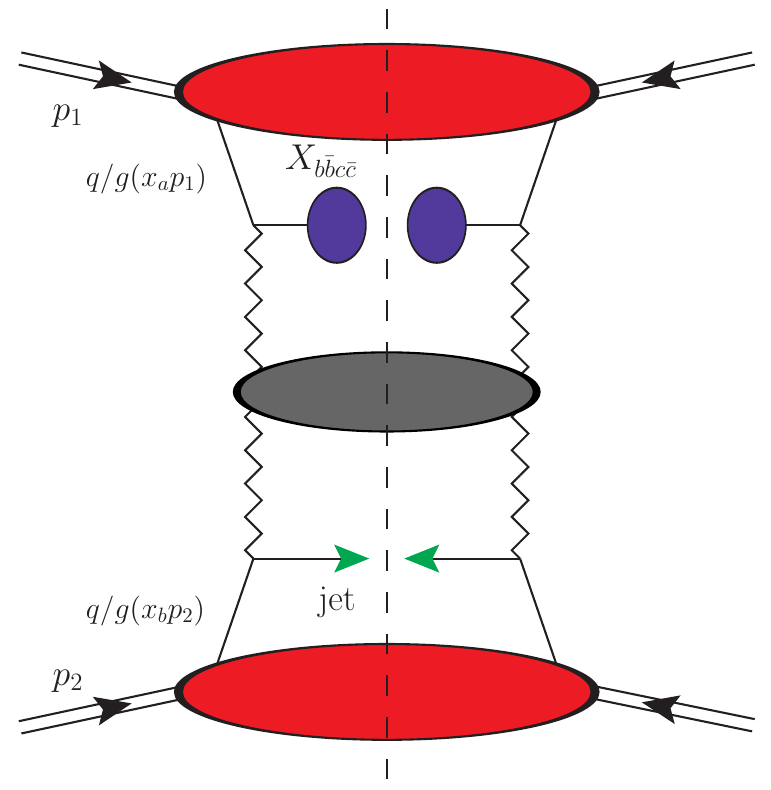}

\caption{Diagrammatic sketch of the semi-inclusive hadroproduction of a $\QXbc$ tetraquark in association with a jet, described within the hybrid high-energy and collinear factorization framework. Red ellipses stand for the collinear parton densities of the incoming protons, and violet ovals denote the fragmentation functions describing the hadronization into the fully heavy tetraquark. 
The jet is depicted by green arrows. Reggeon lines (zigzag) connect the two impact factors to the BFKL Green's function, shown as the central gray ellipse.
Feynman diagrams were produced with {\tt JaxoDraw 2.0}~\cite{Binosi:2008ig}.}
\label{fig:Xbbcc_diagram}
\end{figure*}

We consider the semi-inclusive reaction illustrated in Fig. \ref{fig:Xbbcc_diagram}
\begin{equation}
\label{process}
{\rm p}(p_1)+{\rm p}(p_2)
\to
\QXbc(\kappa_1,y_1)+{\cal X}+{\rm jet}(\kappa_2,y_2)
\,,
\end{equation}
where ${\cal X}$ denotes the unobserved hadronic radiation accompanying the tagged tetraquark and jet.
Predictions are presented for proton-proton collisions at the 14~TeV HL-LHC and at the nominal 100~TeV FCC energy.
We concentrate on semihard configurations in which both tagged objects carry large transverse momenta and are separated by a sizeable rapidity interval, thereby enhancing sensitivity to high-energy logarithmic dynamics.

The calculation is performed within the $\NLLp$ HyF framework~\cite{Celiberto:2015yba,Celiberto:2017ptm,Celiberto:2020wpk}.
It combines NLO collinear ingredients with the resummation of high-energy logarithms at next-to-leading-logarithmic (NLL) accuracy through the NLO Balitsky--Fadin--Kuraev--Lipatov (BFKL) resummation~\cite{Fadin:1975cb,Balitsky:1978ic}.
The `$+$' prescription further retains formally subleading contributions generated by products of NLO corrections to the impact factors.
A detailed account of the HyF formulation for exotic-hadron plus jet production can be found, for instance, in Sec.~III.B of Ref.~\cite{Celiberto:2024beg}.

High-energy factorization and BFKL resummation have been extensively explored in a broad class of semihard reactions.
Representative applications include Mueller--Navelet jets~\cite{Mueller:1986ey,Ducloue:2013hia,Celiberto:2015yba,Celiberto:2016ygs,Celiberto:2017ius,deLeon:2021ecb,Celiberto:2022gji,Baldenegro:2024ndr}, dihadron production~\cite{Celiberto:2016hae,Celiberto:2017ptm,Celiberto:2017ius,Celiberto:2020rxb,Celiberto:2022rfj}, hadron-jet correlations~\cite{Bolognino:2018oth,Celiberto:2020wpk,Celiberto:2020rxb,Celiberto:2022kxx}, forward Higgs channels~\cite{Celiberto:2020tmb,Mohammed:2022gbk,Celiberto:2022zdg}, Drell--Yan production~\cite{Golec-Biernat:2018kem,Celiberto:2018muu}, and heavy-flavor observables~\cite{Bolognino:2019yls,Bolognino:2021mrc,Celiberto:2021dzy,Celiberto:2021fdp,Celiberto:2022dyf,Celiberto:2022grc,Celiberto:2022keu,Celiberto:2024omj,Bulhoes:2025pnp}.
Heavy-flavor tagged final states are particularly appealing in this context because they generally exhibit an enhanced perturbative stability with respect to analogous light-hadron reactions~\cite{Celiberto:2022grc,Bolognino:2018oth,Celiberto:2020wpk}.
Fragmentation-based HyF analyses have already been developed for heavy mesons and quarkonia~\cite{Celiberto:2022dyf,Celiberto:2023fzz,Celiberto:2022keu,Celiberto:2024omj}, and subsequently extended to exotic hadrons, including tetraquarks~\cite{Celiberto:2023rzw,Celiberto:2024beg,Celiberto:2024mab,Celiberto:2025dfe}, pentaquarks~\cite{Celiberto:2025ipt,Celiberto:2026rdk}, and triply heavy baryons~\cite{Celiberto:2025ogy,Celiberto:2026qiz}.
The present analysis brings the all-heavy mixed-flavor $b\bar b c\bar c$ sector into this phenomenological program, with particular emphasis on how its multimodal fragmentation structure affects collider observables and expected production yields.

The phenomenological analysis proceeds from theory control to collider observables and, finally, to production-level observability.
We first quantify the uncertainties associated with the hard-scattering calculation, the fragmentation evolution, and the nonperturbative fragmentation inputs in Sec.~\ref{ssec:uncertainty}.
We then examine the rapidity-separation distributions in Sec.~\ref{ssec:DY_distributions}, which probe how the two microscopic formation mechanisms propagate through the high-energy production dynamics.
Finally, Sec.~\ref{ssec:event_yields} presents integrated rates and event-yield estimates at the HL-LHC and FCC, providing the most direct assessment of the statistical reach available for future searches for the $\QXbc$ signal. 
All numerical predictions are obtained with the {\Jethad} engine~\cite{Celiberto:2020wpk,Celiberto:2022rfj,Celiberto:2023fzz,Celiberto:2024mrq,Celiberto:2024swu,Celiberto:2025csa}.
The proton content is described through the {\tt NNPDF4.0} NLO collinear parton-distribution-function (PDF) set~\cite{NNPDF:2021uiq}, interfaced through the {\tt LHAPDF6} library~\cite{Buckley:2014ana}.

\subsection{Uncertainty quantification}
\label{ssec:uncertainty}

A reliable comparison between the direct-multiquark and scalar-diquark scenarios requires the uncertainty budget to follow the calculation from the hard process down to the fragmentation input.
We therefore separate the theoretical systematics according to their dynamical origin, distinguishing hard-scattering, fragmentation-evolution, hadron-structure, PDF, and numerical effects.

\medskip
\noindent
\textit{Hard-scattering uncertainty (H-MHOU).}
Missing higher-order effects in the HyF calculation are probed through variations of the characteristic hard scales around their natural values.
The corresponding H-MHOU band measures the residual perturbative dependence of the observable beyond the nominal $\NLLp$ accuracy and is evaluated independently of the fragmentation-sector variations.

\medskip
\noindent
\textit{Fragmentation-evolution uncertainty (F-MHOU).}
A second perturbative component originates from the scales defining the starting point of the {\HFNRevo} evolution.
For {\tt TQ4Qhf1.0}, these variations are already embedded in the FF replica ensemble through alternative evolution-ready initial-scale configurations for the direct-multiquark and scalar-diquark modes.
Their propagation to collider observables therefore occurs directly through the corresponding evolved FF replicas.

\medskip
\noindent
\textit{Hadron-structure uncertainty (F-NPWF).}
The nonperturbative modeling of the initial-scale FFs introduces an additional sensitivity to the internal momentum distribution of the tetraquark constituents.
Within the Suzuki-inspired construction, this dependence is monitored through variations of the transverse-momentum parameter $\vqTTa$.
The resulting F-NPWF replicas quantify the response of the phenomenology to changes in the assumed bound-state momentum profile.
We emphasize that the quoted uncertainty bands are evaluated at fixed phenomenological normalization inputs.
In particular, they do not include the independent variation of the scalar-diquark form-factor scale $\Lambda_{\mathcal D}$ discussed in Eq.~\eqref{eq:LambdaD_variation}.
The latter represents an additional source of model dependence, rather than a perturbative or replica-based uncertainty.

\medskip
\noindent
\textit{PDF dependence.}
The proton structure is described through the {\tt NNPDF4.0} NLO set~\cite{NNPDF:2021uiq}.
PDF-induced effects are expected to remain subleading with respect to the hard and fragmentation uncertainties in the semihard kinematics considered here.
We therefore employ the central PDF member throughout the analysis and do not include a separate PDF band in the quoted uncertainty budget.

\medskip
\noindent
\textit{Numerical uncertainty.}
The evaluation of the rapidity-differential observables involves multidimensional phase-space integrations together with the numerical treatment of the high-energy representation implemented in {\Jethad}.
The corresponding integration uncertainty is steadily maintained below the $2\%$ level throughout the phenomenological analysis and therefore remains subleading with respect to the dominant theoretical contributions.
It is nevertheless retained in the construction of the total uncertainty bands.

\medskip

The fragmentation-related F-MHOU and F-NPWF components are thus propagated through the {\tt TQ4Qhf1.0} replica structure itself, while H-MHOU and numerical effects enter at the observable level.
The total theory uncertainty is obtained by combining these independent contributions in quadrature, separately for the upward and downward directions.
The replica-to-parameter correspondence underlying the fragmentation sector is reported in Appendix~\hyperlink{app:A}{A}.

\subsection{$\DY$-distributions}
\label{ssec:DY_distributions}

We first investigate how the multimodal fragmentation structure encoded in the {\tt TQ4Qhf1.0} set propagates to the rapidity separation between the tagged tetraquark and the accompanying jet.
The corresponding observable is the $\DY$-differential rate, with $\DY=y_1-y_2$, obtained by integrating the azimuthally averaged coefficient ${\cal C}_0$ over the accessible transverse-momentum and rapidity phase space at fixed $\DY$,
\begin{equation}
\label{obs:I}
\frac{\drv \sigma(\DY,s)}{\drv \DY}
=
\int_{|\vec\kappa_1|^{\rm min}}^{|\vec\kappa_1|^{\rm max}}
\!\!\drv |\vec\kappa_1|
\int_{|\vec\kappa_2|^{\rm min}}^{|\vec\kappa_2|^{\rm max}}
\!\!\drv |\vec\kappa_2|
\int_{\max(y_1^{\rm min},\,\DY+y_2^{\rm min})}^{\min(y_1^{\rm max},\,\DY+y_2^{\rm max})}
\drv y_1\,
{\cal C}_0^{\rm [res]}
\Bigm|_{y_2=y_1-\DY}
\;,
\end{equation}
where ${\rm [res]}$ specifies the perturbative accuracy adopted in the calculation.
We consider three representative levels of accuracy.
The $\LL$ prediction combines the leading-logarithmic BFKL Green's function with leading-order forward emission functions.
The $\NLLp$ result includes the NLL high-energy kernel together with the NLO corrections to both forward hard factors.
As already mentioned, the superscript `$+$' indicates that, in addition to the terms required by strict NLL/NLO accuracy, the product of the two NLO corrections to tetraquark and jet impact factors is retained.
Finally, $\HENLOp$ denotes the fixed-order high-energy expansion of the resummed result through NLO accuracy, supplemented by the same product of NLO hard-factor corrections.
It therefore provides a natural fixed-order benchmark for assessing genuine high-energy resummation effects.

We impose $40<|\vec\kappa_1|/{\rm GeV}<120$ for the tetraquark and
$50<|\vec\kappa_2|/{\rm GeV}<120$ for the jet, together with
$|y_1|<2.5$ and $|y_2|<4.7$.
These selections keep the tagged system within a semihard regime where the leading-power fragmentation description and the high-energy treatment can be consistently applied.
They also place the calculation in a kinematic domain compatible with the $n_f=5$ ZM-VFNS realization underlying the {\tt TQ4Qhf1.0} determination, where heavy-quark mass effects beyond those already encoded in the initial FFs are expected to enter predominantly through power-suppressed corrections.
The $\DY$ spectra are evaluated in uniform bins of width $\Delta Y_{\rm bin}=0.5$, maintaining a direct connection with prospective experimental analyses.

The perturbative stability of the predictions against variations of the transverse-momentum selection represents an important consistency requirement for the present analysis.
We have therefore performed dedicated checks, not shown here, by varying the lower transverse-momen\-tum bounds entering the $\QXbc$ plus jet configuration.
The resulting spectra preserve a regular hierarchy among the $\LL$, $\NLLp$, and $\HENLOp$ predictions, together with a stable behavior of the corresponding perturbative uncertainty bands.
At the same time, the expected decrease of the absolute rates under progressively harder selections can be naturally traced to the reduction of the available phase space, rather than to a deterioration of the perturbative description.

These findings further support the choice of the semihard kinematic regime adopted throughout our phenomenological analysis.
In particular, the absence of anomalous distortions when the lower transverse-momentum bounds are raised indicates that the predictions are not driven by configurations close to the lower edge of the acceptance, where power-suppressed effects and limitations of the leading-power fragmentation description could become increasingly relevant.
While the cut-variation study does not constitute a direct validation of the ZM-VFNS approximation, it provides an additional phenomenological consistency check of the kinematic domain in which the {\tt TQ4Qhf1.0} FFs are employed.

The same qualitative robustness was previously observed for other rare heavy-flavor final states, notably triply heavy $\Omega_{3c}$ and $\Omega_{3b}$ baryons, under analogous variations of the transverse-momentum selection; see Sec.~V.B of Ref.~\cite{Celiberto:2026qiz}.
The present $\QXbc$ results therefore extend this pattern to all-heavy mixed-flavor tetraquark production and reinforce the overall consistency of the adopted semihard fragmentation setup.

\begin{figure*}[!t]
\centering

   \hspace{0.00cm}
   \includegraphics[scale=0.410,clip]{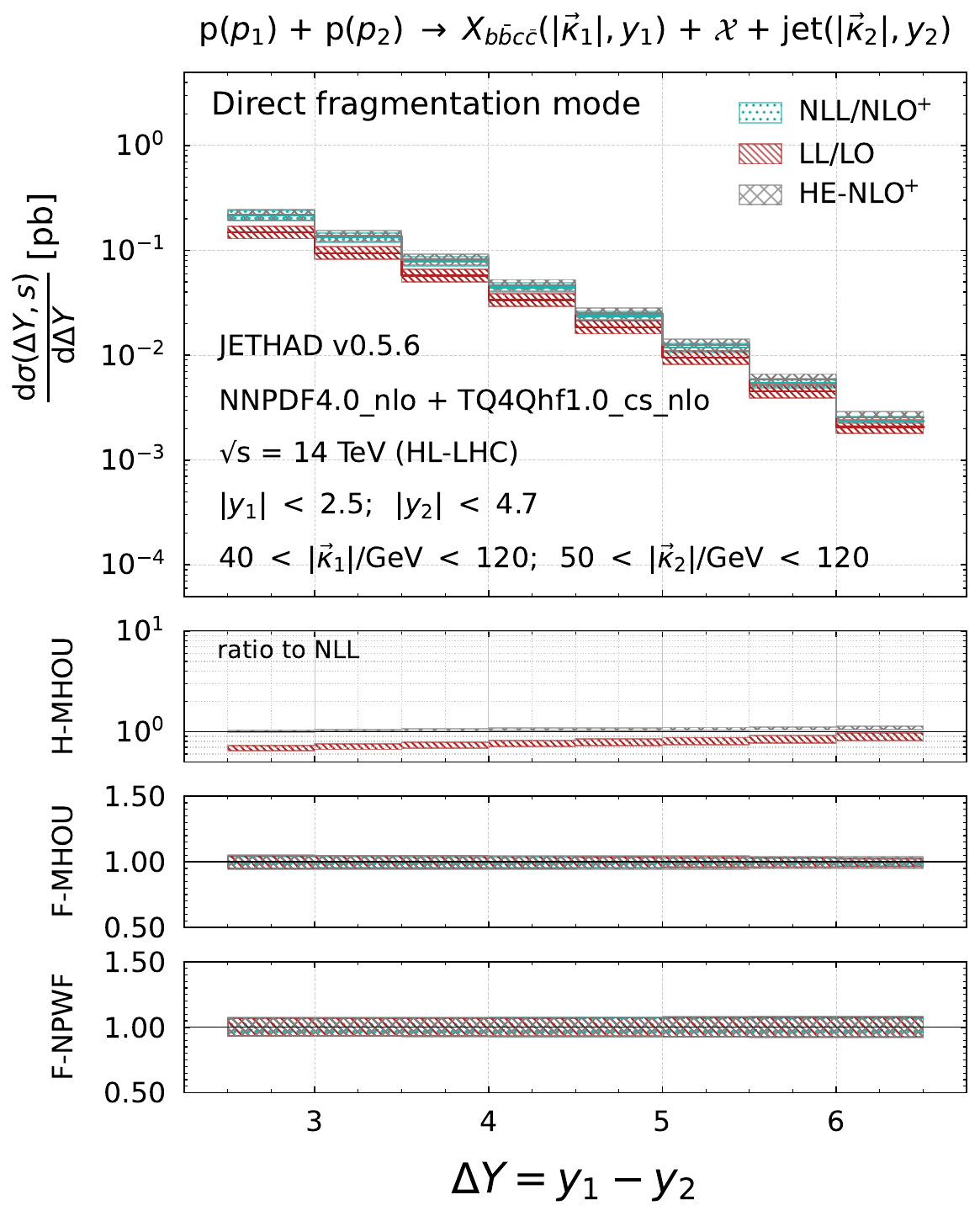}
   \hspace{0.30cm}
   \includegraphics[scale=0.410,clip]{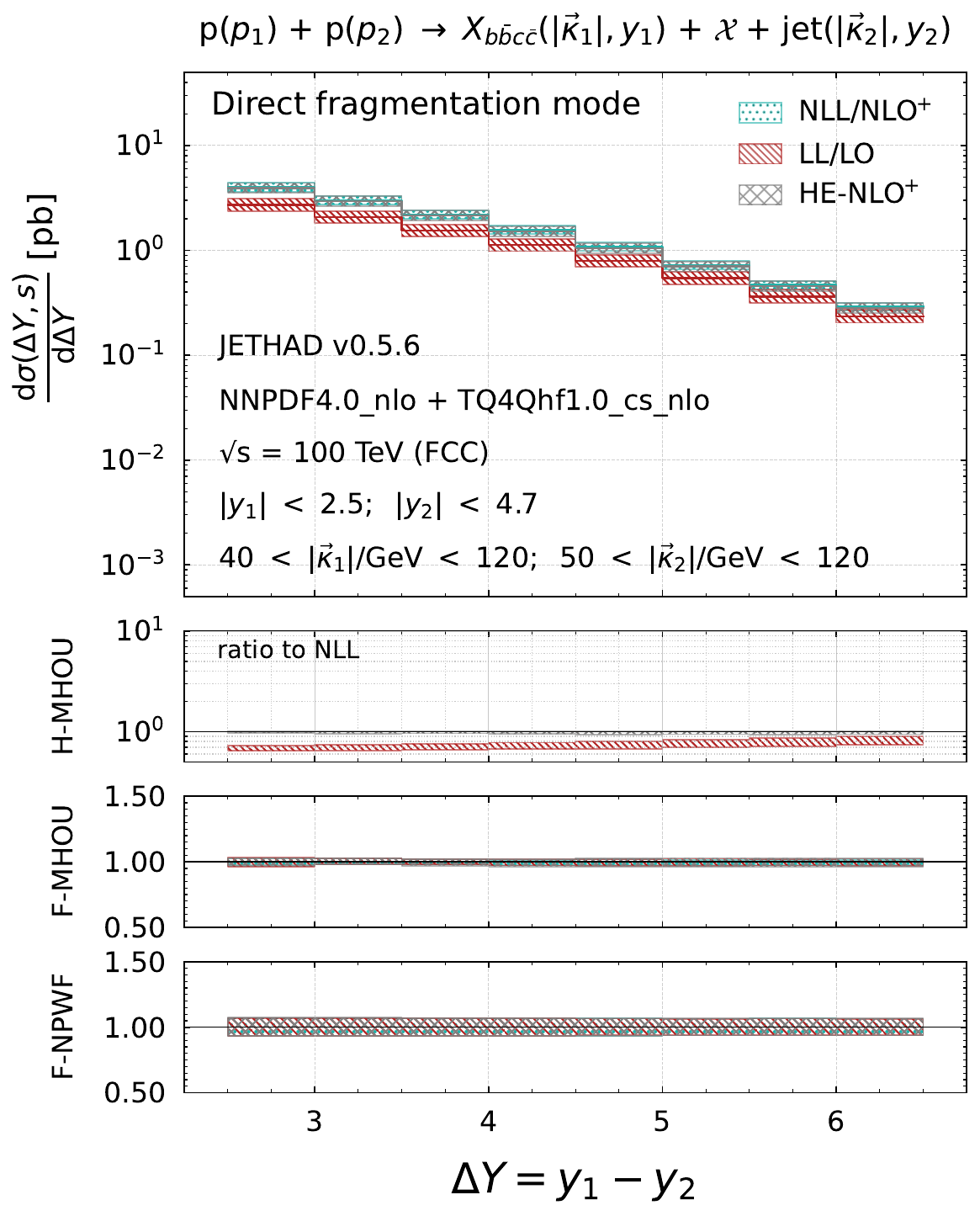}

\caption{Predictions for the $\DY$-distributions of semi-inclusive $\QXbc$ plus jet production in the direct-multiquark fragmentation mode at the 14~TeV HL-LHC (left) and the 100~TeV nominal FCC (right).
Results are shown at $\NLLp$, $\LL$, and $\HENLOp$ accuracy.
The upper panels display the total uncertainty bands, combining H-MHOU, F-MHOU, F-NPWF, and numerical-integration uncertainties.
The lower panels show, from top to bottom: $a)$ the $\LL$- and $\HENLOp$-to-$\NLLp$ ratios, including H-MHOU variations only; $b)$ the F-MHOU uncertainty band; and $c)$ the F-NPWF uncertainty band, with the latter two normalized to their respective central predictions.}

\label{fig:I_Xbc_drt}
\end{figure*}

\begin{figure*}[!t]
\centering

   \hspace{0.00cm}
   \includegraphics[scale=0.410,clip]{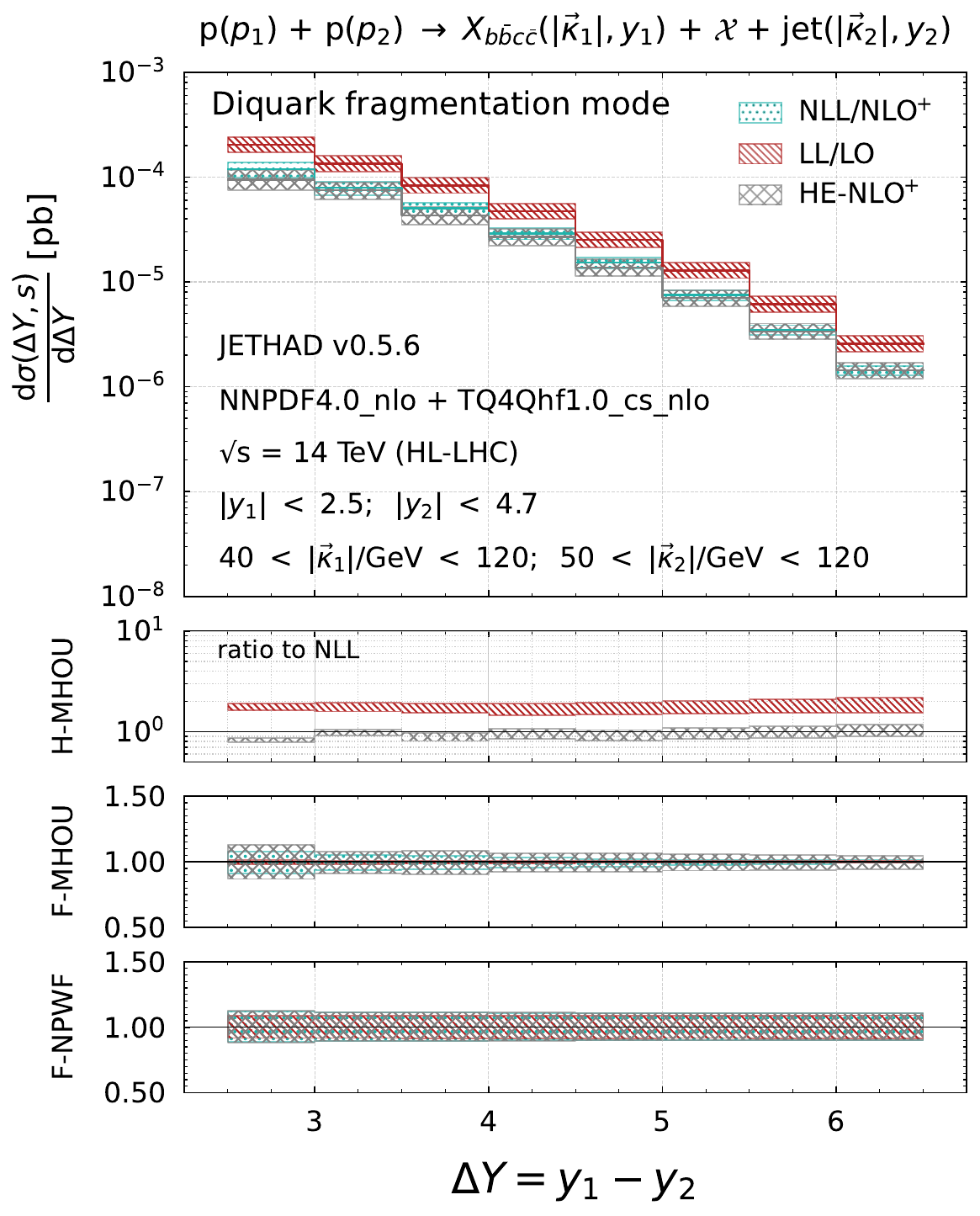}
   \hspace{0.30cm}
   \includegraphics[scale=0.410,clip]{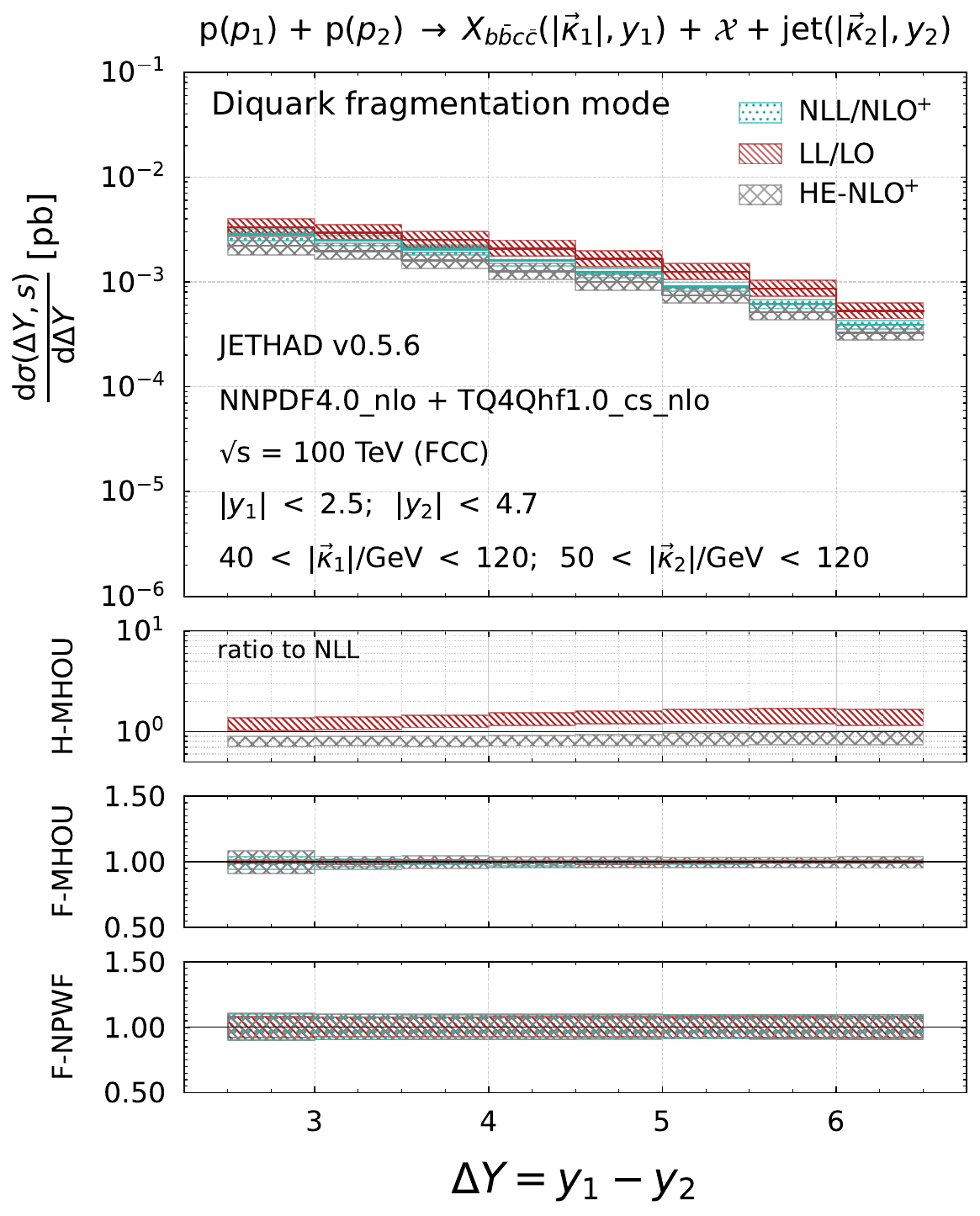}

\caption{Predictions for the $\DY$-distributions of semi-inclusive $\QXbc$ plus jet production in the scalar-diquark fragmentation mode at the 14~TeV HL-LHC (left) and the 100~TeV nominal FCC (right).
Results are shown at $\NLLp$, $\LL$, and $\HENLOp$ accuracy.
The upper panels display the total uncertainty bands, combining H-MHOU, F-MHOU, F-NPWF, and numerical-integration uncertainties.
The lower panels show, from top to bottom: $a)$ the $\LL$- and $\HENLOp$-to-$\NLLp$ ratios, including H-MHOU variations only; $b)$ the F-MHOU uncertainty band; and $c)$ the F-NPWF uncertainty band, with the latter two normalized to their respective central predictions.}

\label{fig:I_Xbc_sdq}
\end{figure*}

Figures~\ref{fig:I_Xbc_drt} and~\ref{fig:I_Xbc_sdq} show the resulting $\DY$ spectra in the direct-multiquark and scalar-diquark modes, respectively, at the 14~TeV HL-LHC and the 100~TeV nominal FCC.
For both fragmentation mechanisms and at both collider energies, the distributions decrease monotonically as the rapidity separation increases.
This behavior results from the competition between the enhancement associated with high-energy evolution and the progressive depletion of longitudinal phase space induced by the convolution with the proton PDFs and tetraquark FFs.
Within the kinematic window considered here, the latter effect prevails and produces the smooth decrease observed throughout the explored $\DY$ range.

Increasing the collision energy to $100$~TeV produces a substantial enhancement of the rates without qualitatively modifying their rapidity dependence.
The increase is already well above one order of magnitude at moderate $\DY$ and becomes progressively more pronounced toward the largest rapidity separations.
This trend reflects the larger partonic luminosities and the wider high-energy phase space available at the FCC, making broadly separated tetraquark--jet configurations considerably more accessible than at the HL-LHC.

A particularly clear manifestation of the multimodal construction emerges from the relative normalization of the two production mechanisms.
Throughout most of the considered $\DY$ range, the direct-multiquark rate exceeds the scalar-diquark contribution by roughly three orders of magnitude.
This hierarchy closely mirrors that observed at the level of the {\tt TQ4Qhf1.0} FFs, where the sequential scalar-diquark mechanism is strongly suppressed with respect to direct fragmentation.
At the same time, the two modes exhibit rather similar $\DY$ slopes.
Hence, despite the markedly different longitudinal-momentum profiles of their underlying FFs, the convolution with the hard production dynamics translates their multimodal structure primarily into a strong normalization hierarchy rather than into radically different rapidity shapes.

The comparison among perturbative accuracies reveals an additional mode-dependent feature.
For direct-multiquark fragmentation, the $\LL$ prediction lies systematically below the $\NLLp$ result, whereas the $\HENLOp$ approximation remains much closer to the latter over the entire rapidity interval.
The scalar-diquark mode displays a different ordering, with the $\LL$ result generally lying above the $\NLLp$ prediction and the $\HENLOp$ contribution remaining below it and comparatively close.
The numerical impact of high-energy corrections is therefore sensitive to the fragmentation input entering the hadronic convolution.
Nevertheless, in both modes the hierarchy among successive approximations evolves smoothly with $\DY$, with no indication of perturbative instabilities in the explored kinematic domain.

The fragmentation-related uncertainty components display a substantially milder dependence on $\DY$.
Both F-MHOU and F-NPWF bands remain relatively narrow and nearly flat over the phenomenologically relevant rapidity interval, showing that these effects predominantly modify the overall normalization rather than the shape of the spectra.
The F-NPWF contribution is particularly smooth, while F-MHOU effects become somewhat more visible in the scalar-diquark channel at the smallest rapidity separations.
More generally, no universal hierarchy among H-MHOU, F-MHOU, and F-NPWF effects emerges across the two fragmentation modes and collider energies, although all components remain under quantitative control.
Numerical-integration uncertainties are steadily maintained below the $2\%$ level.

Overall, the $\DY$ spectra provide a stable differential probe of the two {\tt TQ4Qhf1.0} fragmentation scenarios.
The direct-multiquark and scalar-diquark modes lead to comparable rapidity shapes but sharply separated production rates, while retaining distinct perturbative patterns under high-energy evolution.
This differential analysis establishes the dynamical baseline for the subsequent assessment of experimental observability.
In the following subsection, we integrate the production rates and translate them into expected event yields at the HL-LHC and FCC, adopting a broader symmetric transverse-momentum selection specifically designed to maximize the accessible signal rate.

\subsection{Integrated rates and event yields}
\label{ssec:event_yields}

To complement the differential analysis and directly address the observability of the $\QXbc$ signal, we provide integrated production rates and corresponding event-yield estimates for both the direct-multiquark and scalar-diquark fragmentation mechanisms at the HL-LHC and FCC.
While the $\DY$ distributions probe the underlying production dynamics, event yields translate these predictions into the expected number of produced tetraquarks and therefore provide a more immediate measure of the statistical reach available for future searches.

The expected number of events is obtained from the $\NLLp$ production rate according to
\begin{equation}
\label{eq:event_yields}
N_{\rm events}(s)
=
{\cal L}_{\rm int}
\int_{-6.5}^{6.5}
\drv\DY\,
\frac{\drv\sigma^{\NLLp}(\DY,s)}
     {\drv\DY}
\;,
\end{equation}
where ${\cal L}_{\rm int}$ denotes the integrated luminosity.
The integration covers the adopted rapidity-separation range $|\DY|<6.5$, including both forward and backward configurations.

For the event-yield analysis, we adopt a broader and symmetric transverse-momentum selection,
$30<|\vec\kappa_{1,2}|/{\rm GeV}<120$,
while retaining $|y_1|<2.5$ and $|y_2|<4.7$.
Relative to the asymmetric setup employed for the differential study, this choice enlarges the available phase space and maximizes the integrated production rate, thereby providing a more direct assessment of the production-level observability of the signal while remaining safely within the semihard domain of the calculation.

We consider two complementary luminosity scenarios.
First, a common reference luminosity of
${\cal L}_{\rm int}=355~{\rm fb}^{-1}$,
corresponding to the luminosity delivered to CMS during Run~3~\cite{CMS:2026Run3}, is assigned to both collider energies.
Using the same luminosity at 14 and 100~TeV isolates the enhancement associated with the increase in center-of-mass energy from the HL-LHC to the FCC.
We then adopt the nominal integrated luminosities
\begin{equation}
{\cal L}_{\rm int}^{\rm HL\mbox{-}LHC}=3~{\rm ab}^{-1},
\qquad
{\cal L}_{\rm int}^{\rm FCC}=20~{\rm ab}^{-1},
\end{equation}
following the corresponding HL-LHC and FCC-hh design projections~\cite{Bruning:2020ufq,FCC:2018vvp,Schulte:2025pgs}.
These scenarios provide the most relevant estimates of the production statistics potentially available in future experimental searches.

The uncertainty decomposition follows the same strategy adopted for the $\DY$ distributions.
H-MHOU, F-MHOU, and F-NPWF effects are propagated separately, while the global estimate combines the hard-scattering and fragmentation-related components according to the prescriptions introduced above.
Presenting the isolated uncertainty sources together with their global combination makes it possible to assess not only the central production rates, but also the robustness of the expected statistics against the dominant theoretical systematics.

The resulting integrated cross sections and event yields are reported in Tables~\ref{tab:event-yields-direct-run3} and~\ref{tab:event-yields-diquark-run3} for the common-luminosity benchmark, and in Tables~\ref{tab:event-yields-direct-nominal} and~\ref{tab:event-yields-diquark-nominal} for the nominal HL-LHC and FCC luminosities.
These estimates provide the most direct production-level test of the phenomenological hierarchy identified in the previous subsection and quantify whether the strong preference for the direct-multiquark mechanism still leaves statistically relevant samples for the more suppressed scalar-diquark mode.

We stress that the quoted yields correspond to produced $\QXbc$ states before accounting for decay branching fractions, detector acceptance, reconstruction efficiencies, and experimental backgrounds.
They should therefore be interpreted as a measure of the available production statistics rather than as a detector-level discovery significance.
Within this scope, they provide a direct quantitative link between the microscopic fragmentation scenarios explored in this work and the prospective observability of the $\QXbc$ tetraquark at the HL-LHC, with the FCC offering a substantially enlarged statistical reach.

\begin{table}[!t]
\centering
\begin{minipage}{\linewidth}
\centering
\small
\setlength{\tabcolsep}{4pt}
\renewcommand{\arraystretch}{2.0}
\begin{tabular}{rrrrrr}
\toprule
\multicolumn{6}{c}{\textbf{Direct-multiquark mode (CMS Run~3 luminosity reference, $\mathcal{L}_{\rm int}=355\,\mathrm{fb}^{-1}$)}} \\[4pt]
\midrule
\noalign{\vskip 9pt}
$\sqrt{s}\,[\mathrm{TeV}]$ & $\sigma_{\rm global}\,[\mathrm{pb}]$ & \shortstack{\strut H-MHOU only\\[5pt]\strut $N/10^{6}$} & \shortstack{\strut F-MHOU only\\[5pt]\strut $N/10^{6}$} & \shortstack{\strut F-NPWF only\\[5pt]\strut $N/10^{6}$} & \shortstack{\strut Global\\[5pt]\strut $N/10^{6}$} \\
\midrule
14 (HL-LHC) & $13.68$ & $4.98^{+0.38}_{-0.34}$ & $4.81^{+0.17}_{-0.15}$ & $5.04^{+0.67}_{-0.60}$ & $4.86^{+0.85}_{-0.77}$ \\
\addlinespace[1pt]
100 (FCC) & $118.03$ & $44.54^{+3.80}_{-3.60}$ & $41.44^{+6.14}_{-9.23}$ & $44.98^{+6.53}_{-6.09}$ & $41.90^{+9.81}_{-11.60}$ \\
\bottomrule
\end{tabular}
\caption{NLL/NLO$^+$ predictions for the integrated cross sections and expected yields of semi-inclusive $\QXbc$ plus jet production in the direct-multiquark fragmentation mode (replicas 0--8) at the 14~TeV HL-LHC and the 100~TeV FCC.
A common reference luminosity of $\mathcal{L}_{\rm int}=355\,\mathrm{fb}^{-1}$, corresponding to the luminosity delivered to CMS during Run~3~\cite{CMS:2026Run3},
is assumed, with results integrated over $|\Delta Y|<6.5$.
The kinematic ranges are $30<|\vec{\kappa}_{1,2}|<120~\mathrm{GeV}$, $|y_1|<2.5$, and $|y_2|<4.7$.
H-MHOU variations are evaluated with replica~0 and hard-scale multipliers $\kappa_H=0.5,1,2$.
F-MHOU and F-NPWF uncertainties are obtained from replicas $(0,3,6)$ and $(0,1,2)$, respectively, at $\kappa_H=1$, using their respective means as central values.
Each uncertainty column isolates the indicated source.
The global central prediction averages all active replicas at $\kappa_H=1$.
Its directional H-MHOU widths are averaged over replicas, while F-MHOU and F-NPWF widths are evaluated about their block means, averaged over blocks, and combined in quadrature.
Hard-subprocess and fragmentation uncertainties are then combined in quadrature separately for the upward and downward directions.}
\label{tab:event-yields-direct-run3}
\end{minipage}
\end{table}

The common-luminosity benchmark provides the cleanest starting point for disentangling the effect of the collider energy from that of the accumulated statistics.
At $\sqrt{s}=14$~TeV, the direct-multiquark mode yields an integrated cross section of $13.68$~pb, corresponding to a global estimate of
$N\simeq4.86\times10^{6}$ events for
${\cal L}_{\rm int}=355~{\rm fb}^{-1}$.
At $100$~TeV, the corresponding cross section rises to $118.03$~pb and the yield reaches
$N\simeq4.19\times10^{7}$.
Thus, at fixed luminosity, increasing the center-of-mass energy enhances the direct-multiquark rate by a factor of about $8.6$.

The energy leverage is even stronger for the scalar-diquark mode.
Its integrated cross section increases from $5.62$~fb at the HL-LHC to $85.76$~fb at the FCC, corresponding to approximately
$2.0\times10^{3}$ and
$3.0\times10^{4}$ produced tetraquarks, respectively, for the same $355~{\rm fb}^{-1}$ reference luminosity.
The resulting enhancement factor, about $15$, is therefore substantially larger than in the direct-multiquark case.

This stronger energy dependence can be qualitatively traced back to the softer fragmentation pattern characterizing the scalar-diquark mechanism.
The sequential
$[Q\to{\cal D}]\otimes[{\cal D}\to X]$
formation process populates smaller longitudinal-momentum fractions than the direct-multiquark channel.
For a tetraquark produced with fixed observed transverse momentum, smaller fragmentation fractions imply, on average, a more energetic parent parton and therefore a stronger kinematic penalty from the partonic luminosities at lower collider energy.
Such configurations become comparatively less suppressed when moving from the HL-LHC to the FCC, where the enlarged center-of-mass energy opens a broader partonic phase space and extends the accessible region toward smaller incoming-parton momentum fractions.
The increasing relevance of evolution-generated gluon contributions at higher energies may further reinforce this trend, although a dedicated partonic-channel decomposition would be required to quantify that effect separately.

\begin{table}[!t]
\begin{minipage}{\linewidth}
\centering
\small
\setlength{\tabcolsep}{4pt}
\renewcommand{\arraystretch}{2.0}
\begin{tabular}{rrrrrr}
\toprule
\multicolumn{6}{c}{\textbf{Scalar-diquark mode (CMS Run~3 luminosity reference, $\mathcal{L}_{\rm int}=355\,\mathrm{fb}^{-1}$)}} \\[4pt]
\midrule
\noalign{\vskip 9pt}
$\sqrt{s}\,[\mathrm{TeV}]$ & $\sigma_{\rm global}\,[\mathrm{fb}]$ & \shortstack{\strut H-MHOU only\\[5pt]\strut $N/10^{3}$} & \shortstack{\strut F-MHOU only\\[5pt]\strut $N/10^{3}$} & \shortstack{\strut F-NPWF only\\[5pt]\strut $N/10^{3}$} & \shortstack{\strut Global\\[5pt]\strut $N/10^{3}$} \\
\midrule
14 (HL-LHC) & $5.62$ & $2.00^{+0.29}_{-0.22}$ & $1.97^{+0.12}_{-0.14}$ & $2.07^{+0.39}_{-0.31}$ & $1.99^{+0.90}_{-1.21}$ \\
\addlinespace[1pt]
100 (FCC) & $85.76$ & $30.49^{+1.31}_{-0.86}$ & $29.62^{+0.87}_{-1.05}$ & $31.70^{+5.71}_{-4.50}$ & $30.44^{+7.50}_{-8.68}$ \\
\bottomrule
\end{tabular}
\caption{Same as Table~\ref{tab:event-yields-direct-run3}, but for the scalar-diquark fragmentation mode (replicas 9--17).
H-MHOU variations use replica~9 with $\kappa_H=0.5,1,2$, while F-MHOU and F-NPWF uncertainties are determined from replicas $(9,12,15)$ and $(9,10,11)$, respectively, at $\kappa_H=1$, using their respective means as central values.
Each uncertainty column isolates the indicated source.
The global central prediction averages all active replicas at $\kappa_H=1$.
Its directional H-MHOU widths are averaged over replicas, while F-MHOU and F-NPWF widths are evaluated about their block means, averaged over blocks, and combined in quadrature.
Hard-subprocess and fragmentation uncertainties are then combined in quadrature separately for the upward and downward directions.}
\label{tab:event-yields-diquark-run3}
\end{minipage}
\end{table}

Despite this different energy dependence, the hierarchy between the two microscopic formation scenarios remains very pronounced.
At $14$~TeV, the integrated direct-multiquark cross section exceeds the scalar-diquark one by approximately
$2.4\times10^{3}$, while at $100$~TeV the ratio decreases to about
$1.4\times10^{3}$.
The FCC therefore reduces the quantitative separation between the two mechanisms without altering their ordering.
This behavior closely mirrors the hierarchy already observed at the level of the {\tt TQ4Qhf1.0} FFs and in the $\DY$ distributions, showing that the strong hierarchy in favor of direct-multiquark
production predicted by the fragmentation models survives the convolution with the hard-scattering dynamics and the integration over the experimentally relevant phase space.

The nominal-luminosity scenarios sharpen this picture from the perspective of prospective observability.
At the HL-LHC, with
${\cal L}_{\rm int}=3~{\rm ab}^{-1}$,
the expected global yield reaches
$N\simeq4.11\times10^{7}$
for the direct-multiquark mode.
Even the much more suppressed scalar-diquark mechanism corresponds to approximately
$1.68\times10^{4}$
produced tetraquarks.
At the FCC, combining the higher collision energy with the nominal
${\cal L}_{\rm int}=20~{\rm ab}^{-1}$
leads to
$N\simeq2.36\times10^{9}$
events in the direct channel and
$N\simeq1.72\times10^{6}$
in the scalar-diquark one.
Relative to the nominal HL-LHC scenario, these values correspond to increases by factors of roughly $57$ and $102$, respectively.
The particularly strong gain for the scalar-diquark mode therefore reflects the combined action of the larger FCC luminosity and its stronger intrinsic energy scaling.

\begin{table}[!t]
\centering
\begin{minipage}{\linewidth}
\centering
\small
\setlength{\tabcolsep}{4pt}
\renewcommand{\arraystretch}{2.0}
\begin{tabular}{rrrrrr}
\toprule
\multicolumn{6}{c}{\textbf{Direct-multiquark mode (future-collider nominal luminosities: HL-LHC/FCC)}} \\[4pt]
\midrule
\noalign{\vskip 9pt}
$\sqrt{s}\,[\mathrm{TeV}]$ & $\sigma_{\rm global}\,[\mathrm{pb}]$ & \shortstack{\strut H-MHOU only\\[5pt]\strut $N/10^{7}$} & \shortstack{\strut F-MHOU only\\[5pt]\strut $N/10^{7}$} & \shortstack{\strut F-NPWF only\\[5pt]\strut $N/10^{7}$} & \shortstack{\strut Global\\[5pt]\strut $N/10^{7}$} \\
\midrule
14 (HL-LHC) & $13.68$ & $4.21^{+0.32}_{-0.29}$ & $4.06^{+0.14}_{-0.13}$ & $4.26^{+0.56}_{-0.51}$ & $4.11^{+0.72}_{-0.65}$ \\
\addlinespace[1pt]
100 (FCC) & $118.03$ & $250.93^{+21.38}_{-20.30}$ & $233.47^{+34.57}_{-52.03}$ & $253.40^{+36.76}_{-34.29}$ & $236.06^{+55.28}_{-65.34}$ \\
\bottomrule
\end{tabular}
\caption{Same as Table~\ref{tab:event-yields-direct-run3}, but assuming the nominal integrated luminosities of $\mathcal{L}_{\rm int}=3\,\mathrm{ab}^{-1}$ at the 14~TeV HL-LHC~\cite{Bruning:2020ufq} and $\mathcal{L}_{\rm int}=20\,\mathrm{ab}^{-1}$ at the 100~TeV FCC~\cite{FCC:2018vvp,Schulte:2025pgs}.
All kinematic settings, replica configurations, and uncertainty prescriptions are kept unchanged.}
\label{tab:event-yields-direct-nominal}
\end{minipage}
\end{table}

The uncertainty information reported in the tables is also relevant for assessing the robustness of these estimates.
For direct-multiquark production, the global uncertainty amounts to roughly $15$--$20\%$ at the HL-LHC and reaches the $20$--$30\%$ level at FCC energy.
The scalar-diquark prediction is less constrained at the HL-LHC, where the global replica combination produces a considerably broader interval, whereas at $100$~TeV its relative uncertainty settles at the few-tens-of-percent level.
The isolated H-MHOU, F-MHOU, and F-NPWF columns further show that their relative importance is not universal across energies and formation modes.
Importantly, however, the overall theory uncertainties remain far too small to challenge the orders-of-magnitude hierarchy separating direct-multiquark and scalar-diquark production.

From a microscopic perspective, the event-yield analysis therefore provides a collider-level manifestation of the multimodal fragmentation picture.
The direct-multiquark and scalar-diquark constructions encode distinct assumptions on how the same $\QXbc$ final state is assembled from the fragmenting heavy parton.
Once embedded into the same standard DGLAP evolution, implemented through the threshold-consistent {\HFNRevo} scheme, and subsequently into the same hybrid high-energy and collinear production framework, these different microscopic inputs generate markedly different collider rates.
The comparison is therefore not merely one between alternative normalizations, but between distinct formation hypotheses consistently propagated from the initial-scale FFs to measurable production-level observables.

\begin{table}[!t]
\begin{minipage}{\linewidth}
\centering
\small
\setlength{\tabcolsep}{4pt}
\renewcommand{\arraystretch}{2.0}
\begin{tabular}{rrrrrr}
\toprule
\multicolumn{6}{c}{\textbf{Scalar-diquark mode (future-collider nominal luminosities: HL-LHC/FCC)}} \\[4pt]
\midrule
\noalign{\vskip 9pt}
$\sqrt{s}\,[\mathrm{TeV}]$ & $\sigma_{\rm global}\,[\mathrm{fb}]$ & \shortstack{\strut H-MHOU only\\[5pt]\strut $N/10^{4}$} & \shortstack{\strut F-MHOU only\\[5pt]\strut $N/10^{4}$} & \shortstack{\strut F-NPWF only\\[5pt]\strut $N/10^{4}$} & \shortstack{\strut Global\\[5pt]\strut $N/10^{4}$} \\
\midrule
14 (HL-LHC) & $5.62$ & $1.69^{+0.24}_{-0.19}$ & $1.66^{+0.10}_{-0.12}$ & $1.75^{+0.33}_{-0.26}$ & $1.68^{+0.76}_{-1.02}$ \\
\addlinespace[1pt]
100 (FCC) & $85.76$ & $171.79^{+7.39}_{-4.84}$ & $166.87^{+4.92}_{-5.92}$ & $178.61^{+32.19}_{-25.37}$ & $171.51^{+42.24}_{-48.88}$ \\
\bottomrule
\end{tabular}
\caption{Same as Table~\ref{tab:event-yields-diquark-run3}, but assuming the nominal integrated luminosities of $\mathcal{L}_{\rm int}=3\,\mathrm{ab}^{-1}$ at the 14~TeV HL-LHC~\cite{Bruning:2020ufq} and $\mathcal{L}_{\rm int}=20\,\mathrm{ab}^{-1}$ at the 100~TeV FCC~\cite{FCC:2018vvp,Schulte:2025pgs}. 
All kinematic settings, replica configurations, and uncertainty prescriptions are kept unchanged.}
\label{tab:event-yields-diquark-nominal}
\end{minipage}
\end{table}

These results provide a cautiously encouraging perspective for future searches.
The direct-multiquark scenario yields very large production-level samples in the considered semi-inclusive channel already at the HL-LHC, while the scalar-diquark mode, despite being strongly suppressed dynamically, still gives event counts at the $10^{4}$ level for the nominal HL-LHC luminosity.
At FCC energies, both scenarios benefit dramatically, with even the scalar-diquark mode reaching the million-event regime.
Such numbers do not by themselves establish experimental observability, since decay branching fractions, detector acceptance, reconstruction efficiencies, and backgrounds can substantially reduce the experimentally usable samples.
They nevertheless show that the underlying production rates are not prohibitively small and provide a favorable starting point for dedicated searches and detector-level studies.

Overall, the event-yield analysis reinforces the central phenomenological message of the {\tt TQ4Qhf1.0} framework.
The direct-multiquark mechanism emerges as the strongly favored formation mode at the level of the predicted production rates for the $\QXbc$ tetraquark, whereas the scalar-diquark scenario remains a quantitatively suppressed but potentially accessible alternative.
The HL-LHC already offers encouraging production statistics, while the FCC would provide not only a substantially larger event sample, but also a stronger lever arm for investigating how different microscopic formation mechanisms propagate through fragmentation dynamics into observable heavy-exotic production.

\section{Conclusions and Outlook}
\label{sec:conclusions}
One of the current theoretical challenges in hadronic physics is to understand the production mechanisms of fully heavy tetraquark states in proton-proton collisions, so as to derive realistic yield predictions that can guide experimentalists in their discovery and study. Since the observation of the $X(6900)$ state, considerable effort has been devoted to modeling its production, with existing results indicating that fragmentation is a promising mechanism for producing and studying fully charmed tetraquarks in $pp$ collisions during the high-luminosity phase of the LHC and at the Future Circular Collider. 

Motivated by this perspective, we  extended the theoretical framework to the all-heavy mixed-flavor $\QXbc$ tetraquark and computed the dominant FFs, which are key inputs for a phenomenological investigation of its production at forthcoming hadron colliders. As emphasized in the Introduction, this is a first step toward our ultimate goal: providing realistic predictions for $\QXbc$ yields to assess its discovery prospects at the next generation of hadron colliders.

We presented the {\tt TQ4Qhf1.0} collinear FF set for the  $\QXbc$ state, combining direct-multiquark and sequential scalar-diquark fragmentation modes.
Their initial conditions were evolved through standard DGLAP dynamics within the threshold-aware {\HFNRevo} scheme~\cite{Celiberto:2025euy,Celiberto:2024mex} and a $n_f=5$ realization of the ZM-VFNS. 
We addressed whether distinct microscopic formation mechanisms for the same all-heavy mixed-flavor tetraquark can leave distinguishable signatures in high-energy production.

A clear separation between the two formation pictures emerged already at the fragmentation level.
The direct-multiquark component exhibited a harder longitudinal structure and a substantially larger normalization, whereas scalar-diquark fragmentation populated softer momentum fractions and remained strongly suppressed.
These characteristic differences survived QCD evolution and propagated to the dynamically generated gluon component, showing that the evolved FFs retained information on their microscopic origin.

We then investigated semi-inclusive $\QXbc$ plus jet production at the 14~TeV HL-LHC.
The rapidity-separation distributions showed that the hierarchy between the two fragmentation modes survived the convolution with proton structure and the high-energy hard subprocess.
Predictions at $\LL$, $\NLLp$, and $\HENLOp$ accuracy displayed a regular perturbative behavior, also under variations of the transverse-momentum cuts.
For the adopted model inputs, the direct-multiquark mechanism remained favored by more than three orders of magnitude in the integrated rate.
At the nominal HL-LHC luminosity of $3~{\rm ab}^{-1}$, the corresponding production-level samples reached the $10^{7}$-event range, while the scalar-diquark mode still yielded samples at the $10^{4}$ level in the considered semi-inclusive channel.

These estimates were intentionally interpreted at the production level.
The {\tt TQ4Qhf1.0} FFs were model-driven inputs rather than quantities extracted from data, and branching fractions, detector acceptance, reconstruction efficiencies, and experimental backgrounds were not included.
The predicted rates therefore provided encouraging starting conditions for future searches, but not detector-level observability claims.
Ultimately, data will determine whether either microscopic pattern is realized in nature and, equally importantly, will provide the information needed to refine the fragmentation inputs and the underlying formation models.

The 100~TeV FCC substantially enlarged this prospective sensitivity.
The softer scalar-diquark mode displayed the stronger relative energy enhancement, consistently with its lower-$z$ fragmentation pattern and the broader partonic phase space available at higher collision energy.
For the nominal FCC luminosity of $20~{\rm ab}^{-1}$, the predicted samples reached the billion-event regime for direct fragmentation and the million-event regime for scalar-diquark formation.
The FCC therefore did not alter the hierarchy between the two mechanisms, but would provide a much stronger statistical lever arm for testing their phenomenological consequences.

The multimodal perspective is central to the broader interpretation of these results.
Different microscopic organizations of the same valence content can generate distinct fragmentation patterns, and collider observables may therefore carry information not only on the existence of an exotic state, but also on how it is formed.
This connection can be further enriched by other spin configurations.
Axial-vector tetraquarks, in particular, may exhibit enhanced sensitivity to intrinsic heavy-quark components~\cite{Celiberto:2025vra}, linking exotic production to intrinsic charm and proton structure~\cite{Brodsky:1980pb,Ball:2022qks,Guzzi:2022rca,NNPDF:2023tyk} and to the complementary spectroscopy-structure interplay discussed in Ref.~\cite{Vogt:2024fky}.

Several concrete developments followed naturally from this study.
The fragmentation description can be extended to light-parton channels, while direct applications to single-inclusive $\QXbc$ production will provide a complementary connection with experimentally accessible trans\-verse-momen\-tum spectra.
On the precision side, fixed-order collinear factorization at least at
NLO accuracy will provide an essential benchmark.
A further natural development will be the investigation of large-$z$
soft-gluon (threshold) resummation and of its interplay with heavy-quark
mass and flavor-threshold effects, which are known to play an important
role in heavy-quark fragmentation~\cite{Cacciari:2001cw,Bonino:2023icn,Gaggero:2022hmv,Maltoni:2022bpy,Cacciari:2024kaa,Ghira:2024nkk}.
Finally, dedicated studies of decay modes and branching fractions will be required to connect production-level predictions with realistic signal estimates.

The broader objective is therefore to turn increasingly precise collider measurements into information on the microscopic dynamics of exotic-hadron formation.
The HL-LHC provides the first realistic arena for such tests, while a future FCC could transform them into precision studies.
If successful, this program would make exotic-hadron production more than a search for new resonances: it would become a way of learning, from the way exotic matter is assembled, how the strong force turns quarks and gluons into matter.

\section*{Data availability}
\label{sec:data_availability}
\addcontentsline{toc}{section}{\nameref{sec:data_availability}}

The {\tt TQ4Qhf1.0} collinear FFs~\cite{TQ4Qhf1.0_2026} for all-heavy mixed-flavor
$X_{b\bar b c\bar c}$ tetraquark are publicly available at
\url{https://github.com/FGCeliberto/Collinear_FFs/}.
They are distributed in {\tt LHAPDF} format and include the complete replica ensemble for both the direct-multiquark and scalar-diquark fragmentation modes.

\section*{Acknowledgments}
\label{sec:acknowledgments}
\addcontentsline{toc}{section}{\nameref{sec:acknowledgments}}

We warmly thank Pedro Bicudo for inspiring discussions on the topics addressed in this work.
We also acknowledge the colleagues and participants of the EXCITED QCD 2026 conference in Granada, Spain, for the warm atmosphere and the many fruitful scientific exchanges. VPG would like to thank the
members of the  Departamento de Física y Matemáticas, Universidad de Alcal\'a (UAH), Spain, for their warm hospitality
during the initial phase of this study.
FGC and JVB were supported from the Atracci\'on de Talento Grant n. 2022-T1/TIC-24176 of the Comunidad Aut\'onoma de Madrid, Spain. VPG and JVB were partially financed by the Brazilian
funding agencies CNPq, CAPES (Finance Code 001),  FAPERGS and INCT-FNA (Process No. 408419/2024-5).

\appendix

\counterwithin*{equation}{section}
\renewcommand\theequation{\thesection\arabic{equation}}

\counterwithin*{table}{section}
\renewcommand\thetable{\thesection\arabic{table}}

\hypertarget{app:A}{
\section{Replica structure of the {\tt TQ4Qhf1.0} functions}
}
\label{app:A}

The {\tt TQ4Qhf1.0} replica-like set is divided into two sectors associated with the direct and scalar-diquark production modes.
Replicas $0$--$8$ describe the direct fragmentation configuration, whereas replicas $9$--$17$ encode the diquark-based mechanism.
Inside each sector, replicas are generated through a combined scan of perturbative and nonperturbative sources of uncertainty.
The perturbative component is modeled through fragmentation-scale variations (F-MHOUs), implemented by rescaling the natural initial evolution scales defined in Eqs.~\eqref{eq:muF0_direct} and~\eqref{eq:muF0_diquark} by a multiplicative factor $\kappa_\mu\in[1/2,2]$.
For every perturbative setup, an additional scan is performed on the transverse-momentum parameter $\vqTTa$, which governs the nonperturbative wave-function profile (F-NPWF), using the values $\vqTTa = 65,\,70,\,75~\text{GeV}^2$.
The resulting ensemble therefore associates each replica with a definite pair of perturbative (F-MHOU) and nonperturbative (F-NPWF) inputs.
The baseline setup corresponds to the central-scale configuration ($\kappa_\mu=1$) together with $\vqTTa = 70~\text{GeV}^2$.
The correspondence between replica labels and parameter choices is reported in Table~\ref{tab:TQ4Qhf10_replicas}.

\begin{table}[t]
\centering
\begin{tabular}{|c|c|r|r|}
\toprule
Replica ID & Mode & $\kappa_\mu$ & $\vqTTa$ [GeV$^2$] \\
\midrule
0  & direct  & 1   & 70  \\
1  & direct  & 1   & 65  \\
2  & direct  & 1   & 75  \\
3  & direct  & 0.5 & 70  \\
4  & direct  & 0.5 & 65  \\
5  & direct  & 0.5 & 75  \\
6  & direct  & 2   & 70  \\
7  & direct  & 2   & 65  \\
8  & direct  & 2   & 75  \\
9  & diquark & 1   & 70  \\
10 & diquark & 1   & 65  \\
11 & diquark & 1   & 75  \\
12 & diquark & 0.5 & 70  \\
13 & diquark & 0.5 & 65  \\
14 & diquark & 0.5 & 75  \\
15 & diquark & 2   & 70  \\
16 & diquark & 2   & 65  \\
17 & diquark & 2   & 75  \\
\bottomrule
\end{tabular}
\caption{Organization of the {\tt TQ4Qhf1.0} replica-like ensemble.
The parameter $\kappa_\mu$ parametrizes perturbative energy-scale variations (F-MHOUs), whereas $\vqTTa$ governs the nonperturbative wave-function profile (F-NPWF).}
\label{tab:TQ4Qhf10_replicas}
\end{table}

\clearpage
\bibliographystyle{apsrev}
\bibliography{bibliography}

\end{document}